\documentclass{article}
\usepackage{graphicx}
\usepackage{epstopdf}
\usepackage{cite}
\usepackage{epsfig}
\usepackage{amsfonts}
 \usepackage{amssymb}
\usepackage{amsmath}
\usepackage{slashed}

\usepackage{hyperref}
\usepackage{graphicx}       
\usepackage{dcolumn}        
\usepackage{bm}            		 
\usepackage{amssymb}
\usepackage{amstext}
\usepackage{color} 
\usepackage{transparent}

\date{\today}

\newcommand{\insertplot}[5]{\begin{figure}
 \hfill\hbox to 0.05in{\vbox to #5in{\vfill
 \inputplot{#1}{#4}{#5}}\hfill}
 \hfill\vspace{-.1in}
 \caption{#2}\label{#3}
 \end{figure}}
 \newcommand{\inputplot}[3]{
 \special{ps: plotfile #1}
}
\newcounter{fig}

\newcommand{\ee}{\end{equation}}
\newcommand{\eea}{\end{eqnarray}}
\newcommand{\bea}{\begin{eqnarray}}

\newcommand{\beq}{\begin{equation}}
\newcommand{\eeq}{\end{equation}}

\newcommand{\ze}{\kern 0.05em}

\usepackage{float}

\begin{document}
\begin{center}

{\Large \bf Charged black holes with axionic-type couplings: \\ classes of solutions and dynamical scalarisation}
\vspace{0.8cm}
\\
{Pedro G. S. Fernandes$^{\dagger}$\footnote{pedrogsilvafernandes@tecnico.ulisboa.pt},  
Carlos A. R. Herdeiro$^{\dagger}$\footnote{carlosherdeiro@tecnico.ulisboa.pt},  
Alexandre M. Pombo$^{\ddagger \Diamond}$\footnote{pomboalexandremira@ua.pt},\\
Eugen Radu$^{\ddagger  \Diamond}$\footnote{eugen.radu@ua.pt} and
Nicolas Sanchis-Gual$^{\dagger}$\footnote{nicolas.sanchis@tecnico.ulisboa.pt}
\vspace{0.3cm}
\\
$^{\dagger}${\small Centro de Astrof\'\i sica e Gravita\c c\~ao - CENTRA,} \\ {\small Departamento de F\'\i sica,
Instituto Superior T\'ecnico - IST, Universidade de Lisboa - UL,} \\ {\small Avenida
Rovisco Pais 1, 1049-001, Portugal}
\vspace{0.3cm}
\\
$^{\ddagger }${\small  Center for Research and Development in Mathematics and Applications (CIDMA),
\\
Campus de Santiago, 3810-183 Aveiro, Portugal}
\vspace{0.3cm}
\\
$^{\Diamond}${\small Departamento de F\'\i sica da Universidade de Aveiro,
Campus de Santiago, 3810-183 Aveiro, Portugal}}
\vspace{0.3cm}
\end{center}

 \date{July 2019}

\begin{abstract}   
\par We consider an augmented Einstein-Maxwell-scalar model including an axionic-type coupling between the scalar and electromagnetic field. We study dyonic black hole solutions in this model. For the canonical axionic coupling emerging from high energy physics, all charged black holes have axion hair.  We present their domain of existence and investigate some physical properties. For other axionic-type couplings, two classes of black hole solutions may co-exist in the model: scalar-free Reissner-Nordstr\"om black holes and scalarised black holes. We show that in some region of the parameter space, the scalar-free solutions are unstable. Then, there is non-uniqueness since new  scalarised black hole solutions with the same global charges also exist, which are entropically preferred over the scalar-free solutions and, moreover, emerge dynamically from the instability of the former.  
\end{abstract}

\newpage
%
 \section{Introduction}\label{S1}
%

\par
\par Three types of bosonic fields, each consistent on its own as a classical relativistic field theory, are used in the physical description of Nature: spin 0, 1 and 2 fields. According to the accepted models, the last two fields are realised, in their massless version, as the electromagnetic and gravitational field, whereas the scalar field is realised, in a massive, self-interacting version, as the Higgs boson. Beyond these concrete realisations, it has  been a recurrent speculation in theoretical physics, ranging from quantum gravity to cosmology, that scalar fields play other roles. One particular example is a pseudo-scalar field known as the \textit{axion}. This (yet unobserved) particle was suggested in order to solve the strong CP problem, by Peccei and Quinn~\cite{pecceiquinn77} (see also \cite{PhysRevLett.40.223,PhysRevLett.40.279,AxionsReviewCP}). But it was later understood that besides fulfilling its original purpose, the axion could have far reaching implications in cosmology, namely as a candidate for dark matter \cite{ABBOTT1983133,DINE1983137,PRESKILL1983127,AxionCosmology}. More recently, ultralight axion-like fields have been suggested to arise naturally from  string theory  compactifications \cite{StringAxions1,StringAxions2}, providing yet another possible origin in the context of a fundamental theory. The theoretical soundness of axions motivated experiments, both proposed and conducted,  to detect axionic imprints - see \textit{e.g.} \cite{ExperimentAxionReview,ExperimentAxion1,ExperimentAxion2}.

In the context of strong gravity, namely black hole (BH) spacetimes, the interplay between the gravitational, electromagnetic and non-standard model scalar fields has a long history. If the scalar field is minimally coupled to the electromagnetic and gravitational field, remarkably, it cannot exist as an equilibrium configuration around spherical BHs~\cite{Mayo:1996mv}. This is a realisation of the celebrated ``no-hair" property of BHs - see, $e.g.$~\cite{ScalarHairReview,Volkov:2016ehx}. If non-minimal couplings are allowed, however, hairy BH solutions exist.  In this paper we shall be interested in such BHs when an axionic-type non-minimal coupling between the scalar and the electromagnetic field are considered, leading to charged BHs with axionic-type hair.

Non-minimal couplings between the electromagnetic and a scalar field, and the corresponding BH solutions, have long been considered in the context of, $e.g.$, Kaluza-Klein theory and supergravity~\cite{Gibbons:1987ps,Garfinkle:1990qj}. More recently, new families of such couplings have been  under scrutiny in the context of BH \textit{spontaneous scalarisation}. Spontaneous scalarisation is a strong gravity phase transition. It occurs when two phases (classes of solutions) co-exist, one of them becoming dynamically preferred.  First proposed by Damour and Esposito-Far\`ese in the context of neutron stars \cite{SSNeutronStars}, it was later shown that BHs can undergo spontaneous scalarisation in some models. The most recent interest, for asymptotically flat spacetimes, has been triggered by~\cite{SSGB1,SSGB2,SSGB3}, centered in a class of models dubbed \textit{extended Scalar-Tensor Gauss-Bonnet}. As pointed in \cite{SSChargedBH1}, however, in what concerns BH spontaneous scalarisation, the latter class of models belong to a wider universality class that also contains  \textit{Einstein-Maxwell-Scalar} (EMS) models. EMS models use a non-minimal coupling between the scalar and Maxwell invariant to induce scalarisation, and so, require the presence of electric or magnetic charge. Using EMS models, it has been possible to establish that spontaneous scalarisation of charged BHs occurs dynamically, leading to scalarised, perturbatively stable BHs~\cite{SSChargedBH1,SSChargedBH2,oEMS1,oEMS2,oEMS3}.

In~\cite{EMSdyons} a classification of EMS models was suggested into two main classes, according to the criterion that they admit, or not, the electrovacuum Reissner-Nordstr\"om (RN) BH as a solution. The same can be done for the augmented EMS models, containing also an axionic type non-minimal coupling, that will be the focus here. In the first class, the RN BH is not a solution and the charged BHs all have non-trivial scalar profiles. For the augmented EMS models, this is the case of the canonical axion coupling. BHs in this model have been studied before~\cite{Weinberg,VCardosoAxions,AxionHair1,AxionHair2,AxionHair3,AxionHair4}, but we shall provide a fresh look at these solutions, exhibiting their full domain of existence. In the second class of models, the RN BH is a solution. We shall illustrate them in the augmented EMS models with a particular choice of non-minimal coupling that allows for the spontaneous scalarisation of the RN BH.  We shall construct the scalarised solutions and show they indeed arise dynamically from the evolution of the scalar-free RN BH, when a scalar perturbation is seeded. 

This paper is organised as follows. In Section~\ref{S2} we present the model, obtain the field equations for the ansatz that describes the class of solutions of interest and provide some details on the construction of the solutions. We shall also review the criteria for the class of models wherein spontaneous scalarisation can occur, and describe how the emergence of the scalarised solutions is computed, in linear theory. Section~\ref{S5} describes two physical relations used in assessing the accuracy of the obtained solutions. Section~\ref{S6} describes the effective potential for spherical perturbations, a simple tool that can establish perturbative stability and diagnose possible instabilities. Section~\ref{S7} includes the bulk of our results, including a toy model, and the analysis of the two main examples considered. We close with a discussion of the results obtained.

%
 \section{The model}\label{S2}
%
\par The model herein is a generalisation of the EMS model discussed in~\cite{SSChargedBH1,SSChargedBH2}. It describes a real scalar field $\phi$ minimally coupled to Einstein's gravity and non-minimally coupled to the two relativistic and gauge invariants that can be built from the Maxwell tensor, $F_{\mu\nu}$, in four spacetime dimensions: the Maxwell invariant, $F^{\mu\nu}F_{\mu\nu}$, and the  parity violating invariant, $F_{\mu \nu} \Tilde{F}^{\mu \nu}$. Here, $\Tilde{F}^{\mu \nu}\equiv\epsilon^{\mu \nu \rho \sigma} F_{\rho \sigma}/ (2\sqrt{-g})$ is the Maxwell's tensor Hodge dual, where $\epsilon^{\mu \nu \rho \sigma}$ denotes the Levi-Civita tensor density and $g$ the determinant of the spacetime metric. The generic model is described via the action (hereafter we shall take units with $4\pi G=1=4\pi\epsilon_0$)
\begin{equation}
    \mathcal{S}= \int d ^ { 4 } x \sqrt { - g } \left[R - 2 \partial _ { \mu } \phi \partial ^ { \mu } \phi - f(\phi) F_{\mu \nu} F^{\mu \nu} - h ( \phi ) F_{\mu \nu} \Tilde{F}^{\mu \nu} \right]\ .
\end{equation}
The functions $f(\phi)$ and $h(\phi)$ determine the non-minimal couplings between the scalar field and the electromagnetic field. For  $h(\phi)=0$, the BH solutions have been constructed and discussed extensively in the literature, $e.g.$~\cite{Gibbons:1987ps,Garfinkle:1990qj,SSChargedBH1,SSChargedBH2,EMSdyons}. Here we shall focus on constructing BH solutions, and discussing some of their physical properties, in the case with $h(\phi) \neq 0$. Some previous work in this direction was considered, $e.g.$ in~\cite{Weinberg,VCardosoAxions}.

Depending on the choices of the non-minimal coupling functions $f(\phi),h(\phi)$ two types of models will be considered: models wherein all charged BHs have scalar hair and models wherein charged BHs can be either scalarised or scalar-free. A generic, spherically symmetric line element to describe both the scalar-free and scalarised solutions is
\begin{equation}
\label{ma}
    d s ^ { 2 } = - N ( r ) e ^ { - 2 \delta ( r ) } d t ^ { 2 } + \frac { d r ^ { 2 } } { N ( r ) } + r ^ { 2 } \left( d \theta ^ { 2 } + \sin ^ { 2 } \theta d \varphi ^ { 2 } \right) \ ,
\end{equation}
where $N(r)\equiv 1-2m(r)/r$, and $m(r)$ is the Misner-Sharp mass function~\cite{MisnerSharp}. Spherical symmetry requires the scalar field $\phi(r)$ to have a radial dependence only, and an electromagnetic 4-potential ansatz of the following type, which allows a possible magnetic charge $P$,
\begin{equation}
    A=V(r) dt - P \cos{\theta} d\varphi \ .
    \label{aa}
\end{equation}
Integrating the trivial angular dependence, one obtains the following effective Lagrangian, from which the equations of motion may be derived,
\begin{equation}
    \mathcal{L}_{\rm eff}=-\frac{e^{-\delta} P^2 f(\phi)}{2r^2} + P V' h(\phi)+ \frac{e^{\delta}}{2} r^2 f(\phi) V'^2 + \frac{e^{-\delta} }{2} \left(1-N-rN'-r^2 N \phi'^2 \right) \ ,
\end{equation}
with the prime denoting a radial derivative. Functions $N,\delta,V,\phi$ have radial dependence only; for ease of notation this dependence will be omitted henceforth. 

To write the equations of motion it is convenient to observe the existence of a first integral
\begin{equation}
    V'=-\frac{Q+Ph(\phi)}{r^2 f(\phi)}e^{-\delta} \ ,
\end{equation}
where $Q$ is an integration constant interpreted as the electric charge measured at infinity. Then, the equations of motion read
\begin{equation}
    \delta'=-r\phi'^2 \ ,
\end{equation}
\begin{equation}
    \left(e^{-\delta} r^2 N \phi' \right)' = -\frac{1}{2} e^{\delta} r^2 \dot f(\phi) V'^2 - P \dot h(\phi) V' + \frac{e^{-\delta} P^2 \dot f(\phi)}{2r^2} \ ,
    \label{c4:eq:scalarfield}
\end{equation}
\begin{equation}
    N'=-\frac{(Q+P h(\phi))^2+P^2 f(\phi)^2+f(\phi) r^2 \left(r^2 N \phi'^2+N-1\right)}{f(\phi)r^3} \ ,
\end{equation}
where the dot denotes differentiation with respect to the scalar field, $e.g.$ $\dot f(\phi) \equiv {d f}/{d \phi}$. To solve this set of coupled, non-linear ordinary differential equations, we have to implement suitable boundary conditions for the desired functions and corresponding derivatives. We assume the existence of an event horizon at $r=r_H>0$ and that the solution possesses a power series expansion in $(r-r_H)$
\begin{equation}
    \begin{array} { l } { N ( r ) = N _ { 1 } \left( r - r _ { H } \right) + \ldots, } \qquad { \delta ( r ) = \delta _ { 0 } + \delta _ { 1 } \left( r - r _ { H } \right) + \ldots, } \\ { \phi ( r ) = \phi _ { 0 } + \phi _ { 1 } \left( r - r _ { H } \right) + \ldots, } \qquad { V ( r ) = v _ { 1 } \left( r - r _ { H } \right) + \ldots \ . } \end{array}
    \label{nhe}
\end{equation}
Plugging these expansions in the field equations, the lower order coefficients are determined to be
\begin{align}
\begin{split}
 	 N_1=-\frac{(Q+P h(\phi_0))^2 + P^2 f(\phi_0)^2 - r_H^2 f(\phi_0)}{r_H^3 f(\phi_0)}\ , \qquad
	\delta _1 = -\phi _1 ^{\ze \ze 2}\ze \ze r_H \ , 
\qquad v_1=-\frac{Q+Ph(\phi_0)}{r_H^2 f(\phi_0)}e^{-\delta_0}\ ,
\\
	 \phi_1=-\frac{2P\left[Q+Ph(\phi_0)\right]f(\phi_0) \dot h(\phi_0) + \dot f(\phi_0) \left(P^2f(\phi_0)^2 - \left(Q+Ph(\phi_0)\right)^2 \right)}{2r_H f(\phi_0) \left[\left(Q+Ph(\phi_0) \right)^2 - r_H^2f(\phi_0) + P^2f(\phi_0)^2 \right]}\ . 
	 \end{split}
\end{align}
One observes that only two of the six parameters introduced in the expansions~\eqref{nhe} are independent, which we choose to be $\phi_0$ and $\delta_0$, the remaining being derived from these ones. The solutions in the vicinity of the horizon are determined by these two parameters, together with $(r_H,Q,P)$. Some physical horizon quantities, such as  the Hawking temperature $T_H$, the horizon area $A_H$, the energy density $\rho(r_H)$ and the Kretschmann scalar $K(r_H)$, are then determined by these five parameters as follows: 
\begin{align}
\begin{split}
 	 T_H=\frac{1}{4\pi} N_1 e^{-\delta_0}\ , \qquad
	A_H=4\pi r_H^2\ , \qquad 
	\rho(r_H) = \frac{2\left( P^2 f(\phi_0)^2 + (Q+P h(\phi_0))^2\right)}{r_H^4 f(\phi_0)}\ ,
\\
	 K(r_H)=\frac{4}{r_H^8 f(\phi_0)^2} \bigg\{ 5[Q+Ph(\phi_0)]^4 - 6r_H^2 [Q+Ph(\phi_0)]^2 f(\phi_0) - 6P^2r_H^2 f(\phi_0)^3 + 5P^4f(\phi_0)^4 &\quad \\ + f(\phi_0)^2 \left(10P^2Q^2+3r_H^4+10P^3 h(\phi_0)[2Q+Ph(\phi_0)] \right) \bigg\}\ .\
\end{split}
\end{align}

\par To obtain the boundary conditions at spatial infinity one performs an asymptotic approximation of the solution in the far field. Then the equations of motion yield:
\begin{equation}
    N ( r )=1 - \frac{2M}{r} + \frac { Q ^ { 2 } + P^2 + Q _ { s } ^ { 2 } } { r^2 } + \ldots\ , \qquad \phi(r)=\frac{Q_{s}}{r}+\ldots \, , \qquad \delta(r)=\frac{Q_{s}^{2}}{2 r^{2}}+\ldots \, , \qquad V(r)=\Phi_e+\frac{Q}{r}+\ldots \, , 
\end{equation}
which introduce three new parameters: the scalar charge $Q_s$,  the electrostatic potential difference between the horizon and infinity $\Phi_e$ and the ADM mass $M$. From these asymptotic expansions one collects a set of 8 independent parameters, therefore: $(r_H,Q,P,\phi_0,\delta_0,Q_s,\Phi_e,M)$. As we shall see below, the full integration of the field equations relates these parameters, and, for each choice of the coupling functions, the solutions of interest actually form a family of solutions with only three (continuous) parameters, typically taken to be the global charges $(M,P,Q)$, but possible with non-uniqueness.

For later use we collect the following results and definitions:
\begin{equation}
    F_{\mu \nu}F^{\mu \nu} = \frac{2\left[P^2 f(\phi)^2-(Q+P h(\phi))^2 \right]}{r^4f(\phi)^2}\ ,\, \qquad F_{\mu \nu}\Tilde{F}^{\mu \nu} = \frac{4P(Q+P h(\phi))}{r^4 f(\phi)} \ , \qquad \beta \equiv \frac{P}{Q} \ ,
\end{equation}
\begin{equation}
    q\equiv \frac{\sqrt{Q^2+P^2}}{M}\ ,\, \qquad a_H\equiv \frac{A_H}{16\pi M^2} \ , \, \qquad t_H\ \equiv 8\pi M T_H\ ,
\end{equation}
where $q$ is the \textit{reduced} dyonic charge, $a_H$ is the \textit{reduced} horizon area and $t_H$ the \textit{reduced} BH temperature. These reduced quantities are convenient because they are invariant under the scaling symmetry
\begin{equation}
    r \to \lambda r, \qquad \xi \to \lambda \xi \ ,
\end{equation}
where $\xi$ represents any of the global charges of the model, while $f(\phi)$ and $h(\phi)$ remain unchanged.

\subsection{Models with spontaneous scalarisation}
Depending on the choice of the functions $f(\phi)$ and $h(\phi)$ the model may accommodate either only BH solutions with a non-trivial scalar field profile, or both scalar-free and scalarised charged BHs. In the latter case, the scalar-free BH solutions may become unstable, within some region of the parameter space, dynamically evolving to the scalarised solutions. This is the phenomenon of spontaneous scalarisation of charged BHs~\cite{SSChargedBH1,SSChargedBH2}. In this subsection we identify the conditions for scalarisation to occur and study its onset, at linear level.

%
\subsubsection{Spontaneous scalarisation conditions}\label{S3}

\par  In order for spontaneous scalarisation of charged BHs to occur, the coupling functions (and their derivatives) must satisfy two key conditions~\cite{SSChargedBH1,SSChargedBH2,SSGB1,SSGB2,SSGB3}. These read as follows:
\begin{enumerate}
      
    \item The system must accommodate a scalar-free solution. Since the Klein-Gordon equation of motion is
    \begin{equation}
        \Box \phi = \frac{\dot f(\phi) F_{\mu \nu}F^{\mu \nu} + \dot h(\phi) F_{\mu \nu}\Tilde{F}^{\mu \nu}}{4} \ ,
        \label{c4:eq:KGeq}
    \end{equation}
   the existence of a scalar-free solution requires (for both $Q,P$ non-zero)
    \begin{equation}
        \dot f(0)=0\ , \qquad \dot h(0)=0 \ .
    \end{equation}
    
    \item Spontaneous scalarisation occurs if the scalar-free solution is unstable against scalar perturbations $\delta \phi$. These obey (neglecting second order terms)
    \begin{equation}
        (\Box - \mu_{\rm eff}^2) \delta \phi = 0 \ ,
    \end{equation}
   where\footnote{For simplicity we have assumed $ f(0)=1$ in ~\eqref{c4:eq:mueff}.}
    \begin{equation}
    \small
        \mu_{\rm eff}^2=\frac{\ddot f(0) F_{\mu \nu}F^{\mu \nu}|_{\phi=0} + \ddot h(0) F_{\mu \nu}\Tilde{F}^{\mu \nu}|_{\phi=0}}{4}=\frac{\ddot f(0) (P^2-(Q+P h(0))^2) + 2\ddot h(0) P(Q+Ph(0))}{2r^4} \ .
        \label{c4:eq:mueff}
    \end{equation}
    The instability occurs if the mass is tachyonic, $i.e.$, $\mu_{\rm eff}^2<0$. This constrains the second derivatives of the coupling functions.
\end{enumerate}
Besides these conditions, one may require, for convenience, that Maxwell's theory should be recovered near spatial infinity. Thus
    \begin{equation}
        f(0)=1 \ .
    \end{equation}
    No condition is imposed on the coupling function $h(\phi)$, since the term $F_{\mu \nu}\Tilde{F}^{\mu \nu}$ is  topological, and hence non-dynamical when the corresponding scalar coupling becomes constant.

%
\subsubsection{Bifurcation of solutions: the existence line} \label{S4}
%
\par Let us now consider the onset of spontaneous scalarisation. We assume that the model under consideration admits the dyonic RN BH of Einstein-Maxwell theory as  the scalar-free solution, that is~\eqref{ma}-\eqref{aa} with
\begin{equation}
\delta=0 \ , \qquad N(r)=1-\frac{2M}{r}+\frac{Q^2+P^2}{r^2} \ , \qquad V(r)=\frac{Q}{r}\ .
\end{equation}
The scalarisation phenomenon is assessed by considering scalar perturbations of the RN solution within the considered model.  Following~\cite{SSChargedBH1,SSChargedBH2}, we take a spherical harmonics decomposition of the scalar field perturbation:
\begin{equation}
    \delta \phi = \sum_{\ell,\mathbf{m}} Y_{\ell,\mathbf{m}}(\theta,\varphi) U_\ell(r) \ .
\end{equation}
With this ansatz, the scalar field equation of motion \eqref{c4:eq:KGeq} simplifies to
\begin{equation}
\label{sfep}
    \frac { e ^ { \delta } } { r ^ { 2 } } \left( \frac { r ^ { 2 } N } { e ^ { \delta } } U _ { \ell } ^ { \prime } \right) ^ { \prime } - \left[ \frac { \ell ( \ell + 1 ) } { r ^ { 2 } } + \mu _ { \mathrm { eff } } ^ { 2 } \right] U _ { \ell } = 0 \ .
\end{equation}
Once the coupling functions are fully fixed, solving~\eqref{sfep} is an eigenvalue problem: for a given $\ell$, requiring an asymptotically vanishing, regular at the horizon, smooth scalar field, a discrete set of BHs solutions are selected, $i.e.$ a discrete set of RN solutions, each with a certain reduced charge $q$. These are the \textit{bifurcation points} from the scalar-free solution. They are labelled by an integer $n\in \mathbb{N}_0$; $n=0$ is the fundamental mode, whereas $n\geqslant 1$ are excited states (overtones). The RN solutions with a smaller (larger) $q$ than that of the bifurcation point are stable (unstable) against the corresponding scalar perturbation. In particular, the first bifurcation point, $i.e.$, the one with the smallest $q$, which corresponds to the mode $\ell=0$ and $n=0$, marks the onset of the scalarisation instability. Only RN BHs with $q$ smaller than the first bifurcation point are stable against any sort of scalar perturbation.

\begin{figure}[ht!]
    \centering
    \includegraphics[scale=0.85]{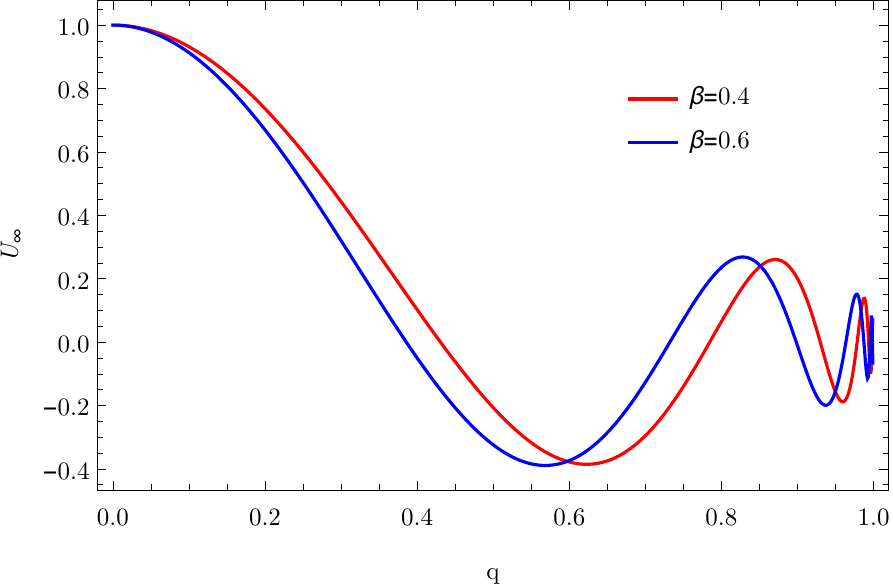}\hfill
    \includegraphics[scale=0.85]{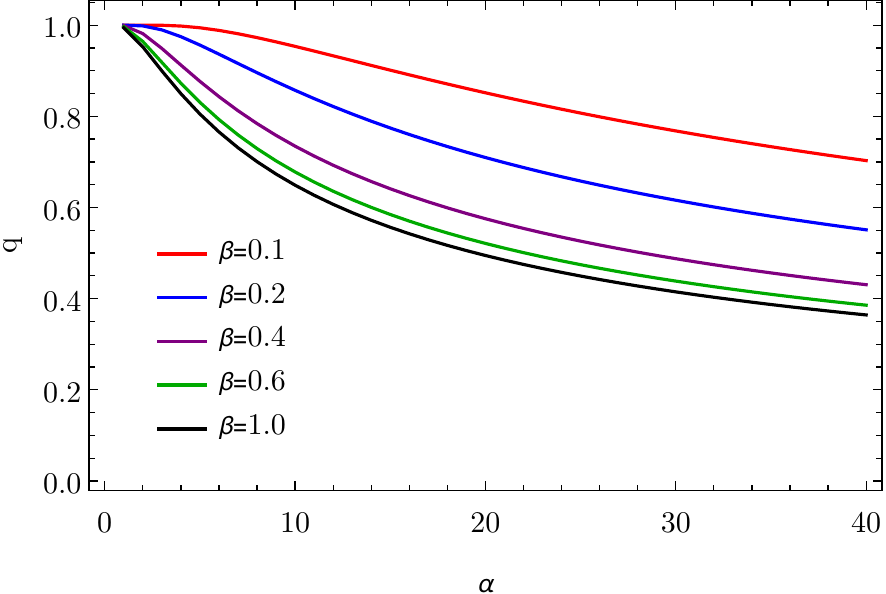}
    \caption{(Left panel) $U_\infty$ as a function of $q$ for two values of $\beta$  and $\alpha=40$. An infinite set of configurations with $U_\infty=0$ exist, labelled by $n$, the number of nodes of $U(r)$. The first configuration for which $U_\infty=0$ is labelled by $n=0$ and so on. (Right panel) Existence line for a sample of $\beta$ values.}
    \label{c4:fig:ExistenceLegendre}
\end{figure}

At each bifurcation point, a new family of (fully non-linear) scalarised BH solutions emerges from the RN family, as static solutions of the equations of motion of the full model. In this paper we shall consider only the first bifurcation point and the corresponding new family of spherically symmetric scalarised BHs that bifurcate from the RN family. To be concrete, let us take $f(\phi)=1$ and $h(\phi)=-\alpha \phi^2$ (with $\alpha>0$), a case we shall consider in detail later. Then $\mu_{\rm eff}^2=-2\alpha Q P/r^4$, which confirms the tachyonic instability.\footnote{In fact, this linear analysis is true for any function $h(\phi)$ that respects $h(0)=0$ and its Taylor expansion possesses a quadratic term of the form $-\alpha \phi^2$.} For $\ell=0$, one finds the following exact solution

\begin{equation}
    U(r)=LP_u\left[ - \frac{(Q^2+P^2)(r-2r_H)+r_H^2 r}{r(Q^2+P^2-r_H^2)}\right], \qquad {\rm where} \qquad u\equiv\frac{1}{2}\left(\sqrt{1-\frac{8\alpha \beta}{1+\beta^2}}-1\right) \ ,
\end{equation}
where $LP_u$ is a Legendre function. The function $U(r)$ approaches a constant \textit{non-zero} value as $r\to \infty$
\begin{equation}
    U(r) \to U_\infty = LP_u\left[ \frac{1+\sqrt{1-q^2}}{1+\sqrt{1-q^2} - q^2} \right] + \mathcal{O}\left(\frac{1}{r}\right)\ ;
\end{equation}
thus, finding the $\ell=0$ unstable mode of the RN BH amounts to a study of the zeros of the Legendre function -- \textit{cf.} Fig. \ref{c4:fig:ExistenceLegendre} (left panel). Bifurcation requires $\alpha$ above a minimum value given by $\alpha_{\rm min} = (1+\beta^2)/(8\beta)$. Some illustrative values can be found in Table \ref{c4:tab:amin}.

\begin{table}[ht!]
\centering
    \begin{tabular}{|c|c|c|c|c|c|c|c|c|c|c|}
        $\beta$ & $0.01$  & $0.1$  & $0.2$  & $0.3$ & $0.4$  & $0.5$  & $0.6$ & $1$ & $2$ & $4$\\ \hline
		$\alpha_{min}$  & $12.5013$ & $1.2625$ & $0.65$  & $0.454167$ & $0.3625$  & $0.3125$ & $0.28333$ & $0.25$ & $0.3125$ & $0.53125$
	\end{tabular}
	\caption{Minimum value of $\alpha$ for bifurcation from a dyonic RN BH for several values of $\beta$. Observe the values of $\alpha_{min}$ are the same for $\beta$ and $1/\beta$, by electromagnetic duality.}
	\label{c4:tab:amin}
\end{table}

\par For $\beta\leq 1$, a smaller $\beta$ requires a larger $\alpha$  for bifurcation; and bifurcation occurs for a larger $q$, in contrast to the dyonic case~\cite{EMSdyons}. Observe that for $P=0$, the Legendre function trivialises, since $F\tilde{F}$ vanishes. Varying $\alpha$, the set of bifurcation points constitutes the \textit{existence line}, $i.e.$, the set of RN configurations for which the zero mode exists. This line is shown in Fig.~\ref{c4:fig:ExistenceLegendre} (right panel) for some illustrative values of $\beta$ in a $q$ $vs.$ $\alpha$ plot.

%
 \section{Physical relations}\label{S5}
%
\par Let us now briefly consider two physical relations that, besides their physical content, are used to test the numerical accuracy of the solutions found numerically. These are a Smarr-type law and a virial-type relation.
%
%
\subsection{A Smarr-type law}
\par The Smarr law \cite{4lawsBH,Smarr} provides a relation between the total mass of the spacetime and other measurable quantities, like the horizon temperature and area. Its information complements that of the equations of motion, making it an interesting test to assess the accuracy of BH solutions obtained numerically. 

The Smarr law can be obtained via the integral mass formula, that for our model reads
\begin{equation}
    M=\frac{1}{2}T_H A_H - \frac{1}{16\pi} \int_V (2T^b_a - T \delta^b_a) k^a d\Sigma_b \ ,
\end{equation}
where $k^a$ is the Killing vector field associated to staticity and $T$ is the trace of the energy-momentum tensor. The full energy-momentum tensor is
\begin{equation}
    T_{\mu \nu} = 4 \left[ f(\phi) \left( F_{\mu \alpha} F_{\nu}^\alpha - \frac{1}{4} g_{\mu \nu} F_{\alpha \beta} F^{\alpha \beta} \right) +  \partial_\mu \phi \partial_\nu \phi - \frac{1}{2} g_{\mu \nu} \partial_\alpha \phi \partial^\alpha \phi \right]\ ,
\end{equation}
which is insensitive to the $F_{\mu \nu}\Tilde{F}^{\mu \nu}$ term in the action, as the latter is  a topological invariant. One can thus arrive at the Smarr-type law
\begin{equation}
    M=\frac{1}{2}T_H A_H + \Phi_e Q + \Phi_m P + U \ ,
\end{equation}
where we refer to 
\begin{equation}
    \Phi_e \equiv \int_{r_{H}}^{\infty} dr \left(\frac{Q+Ph(\phi)}{r^2 f(\phi)}e^{-\delta}\right) \equiv -\int_{r_{H}}^{\infty} d r V^\prime,\ , \qquad \Phi_m \equiv  \int_{r_{H}}^{\infty} d r \left(e^{-\delta} f(\phi) \frac{P}{r^2} \right) \equiv - \int_{r_{H}}^{\infty} d r V_m^\prime \ ,
\end{equation}
as the electrostatic and magnetostatic potential differences, respectively, and
\begin{equation}
    U \equiv  - \int_{r_H}^{\infty} dr V^\prime P h(\phi) = \frac{1}{16 \pi} \int F_{\mu \nu} \Tilde{F}^{\mu \nu} h(\phi) d^3x \ ,
\end{equation}
is interpreted as the axion-related electromagnetic energy. A non-linear Smarr-type relation can also be established for this family of models, following~\cite{SSChargedBH1}, which reads
\begin{equation}
    M^2+Q_s^2+U^2=Q^2+P^2+\frac{1}{4}A_H^2T_H^2 \ .
\end{equation}

\subsection{A virial-type relation}
\par Scaling arguments, initiated by the work of Derrick~\cite{Derrick:1964ww}, are a powerful tool to establish no-go theorems for solitonic solutions (see $e.g.$~\cite{solitonsEMS}),  no-hair theorems for BH solutions \cite{ScalarHairReview}, as well as to provide a physical relation that must be obeyed by solutions of a given model. These relations are generalisations of the canonical virial theorem, that states an energy balance, and are often described as virial relations. They are typically independent from the equations of motion; thus, again, they are useful in assessing the accuracy of numerically generated solutions.

Consider the effective action
\begin{equation}
    \mathcal{S}_{\rm eff} = \int_{r_H}^{\infty} dr \mathcal{L}_{\rm eff}\ ,
\end{equation}
and assume that a charged BH solution with scalar hair exists, described by the functions $\phi(r),\delta(r),V(r),N(r)$, with suitable boundary conditions at the event horizon and at infinity. Next, consider the 1-parameter family of configurations described by the scaled functions 
\begin{equation}
    F_\lambda(r) \equiv F(r_H+\lambda(r-r_H))\ ,
\end{equation}
with $F \in \{\phi,\delta,V,N\}$. If the initial configuration was indeed a solution, then the effective action for the scaled configurations must possess a critical point at $\lambda=1$:  $\left({dS_{eff}^\lambda}/{d\lambda}\right)_{\lambda=1}=0$. From this condition one obtains the virial-type relation 
\begin{equation}
    \int_{r_{H}}^{\infty} d r\left\{ e^{-\delta} r^{2} {\phi^\prime}^2\left[1-\frac{r_{H}}{r}\left(1+N\right)\right]\right\} = \Phi_e Q + \Phi_m P + U + \int_{r_H}^{\infty} d r \left\{\frac{2r_H}{r}\left[V^\prime Q+ V_m^\prime P + V^\prime P h(\phi) \right]\right\}\ .
\end{equation}
\par One can show that the left hand side integrand is strictly positive. Thus, the virial identity shows that a nontrivial scalar field requires a nonzero electric/magnetic charge so that the right hand side is nonzero. As an immediate corollary, neutral BHs cannot be hairy in this model. 

%
\section{Effective potential for spherical perturbations}\label{S6}
%
\par Let us also introduce a diagnosis analysis of perturbative stability, against spherical perturbations, that shall be applied to the solutions derived and discussed in the next section. 

Following a standard technique, see $e.g.$ \cite{SSChargedBH2}, we consider spherically symmetric, linear perturbations of an equilibrium solution, keeping the metric ansatz, but allowing the functions $N,\delta,\phi,V$ to depend on $t$ as well as on $r$:
\begin{equation}
    ds^2=- \tilde N(r,t)e^{-2 \tilde \delta(r,t)} dt^2+\frac{dr^2}{\tilde N(r,t)}+r^2(d\theta^2+\sin^2 \theta d\varphi^2) \ ,
 \qquad A= \tilde V(r,t) dt-P\cos\theta d\varphi \ , \qquad \phi=\tilde \phi(r,t) \ .
\end{equation}
The time dependence enters as a Fourier mode with frequency $\Omega$, for each of these functions:
\begin{eqnarray}
&&
 \tilde N(r,t)=N(r)+\epsilon N_1(r)e^{-i \Omega t}\ , \qquad  \tilde \delta(r,t)=\delta(r)+\epsilon \delta_1(r)e^{-i \Omega t} \ , 
\\
&&
\nonumber
 \tilde \phi(r,t)=\phi(r)+\epsilon \phi_1(r)e^{-i \Omega t}\ , \qquad \tilde V(r,t)=V(r)+\epsilon V_1(r)e^{-i \Omega t}\ .
\end{eqnarray}
From the linearised field equations around the background solution, the metric perturbations and $V_1(r)$ can be expressed in terms of the scalar field perturbation,
\begin{eqnarray}
N_1=-2r N\phi' \phi_1 \ , \qquad \delta_1'=-2 r\phi' \phi_1' \ , \qquad V_1'= -V'\left( \delta_1 + \frac{\dot f(\phi)}{f(\phi)}\phi_1 \right)+ \frac{e^{-\delta} P \dot h(\phi)}{f(\phi) r^2} \phi_1 \ ,
\end{eqnarray}
thus yielding a single perturbation equation for $\phi_1$. Introducing a new variable $\Psi(r)=r\phi_1$, the scalar-field equation of motion may be written as
\begin{equation}
    \left(N e^{-\delta}\right)^2 \Psi'' + N e^{-\delta}\left(N e^{-\delta}\right)' \Psi' + \left(\Omega^2 - U_\Omega\right) \Psi = 0 \ ,
\end{equation}
which, by introducing the 'tortoise' coordinate $x$ as $dx/dr=e^{\delta}/N$ \cite{qnmcardoso}, can be written in the standard one-dimensional Schr\"odinger-like form:
\begin{equation}
    -\frac{d^2 }{dx^2}\Psi+U_{\Omega} \Psi=\Omega^2 \Psi \ .
    \label{c4:eq:stab-shrod}
\end{equation}
The effective potential that describes spherical perturbations $U_{\Omega}$ is defined as:
\begin{equation}
    U_\Omega = U_0 + P U_1 + P^2 U_2 \ ,
\end{equation}
with
\begin{align}
    \begin{split}
        U_0&=\frac{e^{-2\delta}N}{r^2} \left\{ 1 - N -2r^2\phi'^2 - \frac{Q^2}{r^2 f(\phi)} \mathcal{U} \right\} \ , \\
        U_1&=\frac{e^{-2\delta}N Q}{r^4 f(\phi)} \left\{ \ddot h(\phi) + 4r\dot h(\phi)\phi' - 2h(\phi) \mathcal{U} \right\} \ , \\
        U_2&=\frac{e^{-2\delta}N}{r^4} \left\{ \frac{\ddot f(\phi)}{2} +2r\phi' \dot f(\phi) - f(\phi)(1-2r^2\phi'^2) + \right. \\ &\quad \left.  \frac{1}{f(\phi)} \left[ \ddot h(\phi) h(\phi) - \dot h(\phi)^2 + 4r\phi'\dot h(\phi)h(\phi) - h(\phi)^2 \mathcal{U}\right] \right\} \ , \\
 {\rm where} \ \        \mathcal{U}&\equiv1-2r^2\phi'^2 + \frac{\ddot f(\phi)}{2 f(\phi)} + 2r\phi' \frac{\dot f(\phi)}{f(\phi)}-\left(\frac{\dot f(\phi)}{f(\phi)}\right)^2 \ .
    \end{split}
\end{align}
An unstable mode would have $\Omega^2<0$, which for the asympotic boundary conditions of our model is a bound state. It follows from a standard result in quantum mechanics (see $e.g.$~\cite{Messiah:1961}), however, that eq. (\ref{c4:eq:stab-shrod}) has no bound states if $U_\Omega$ is everywhere larger than the lowest of its two asymptotic values, $i.e.$, if it is positive in our case.\footnote{A simple proof is as follows. Write Eq. (\ref{c4:eq:stab-shrod}) in the equivalent form \begin{eqnarray}
 \frac{d }{dx}\left(\Psi \frac{d \Psi}{dx}\right)=\left(\frac{d \Psi}{dx}\right)^2+(U_{\Omega}  -\Omega^2) \Psi^2~.
\end{eqnarray}
After integrating from the horizon to infinity it follows that
\begin{eqnarray}
\int_{-\infty}^{\infty}dx \left[ \left(\frac{d \Psi}{dx}\right)^2+  U_{\Omega} \Psi^2 \right]=\Omega^2  \int_{-\infty}^{\infty}dx \Psi^2
\end{eqnarray}
which for $ U_{\Omega} >0$ implies $\Omega^2 >0$. 
} Thus an everywhere positive effective potential proofs mode stability against spherical perturbations.

We remark that the existence of a region of negative potential is a necessary but not sufficient condition for instabilities to be present. In fact, for the fundamental, spherically symmetric scalarised solutions in~\cite{SSChargedBH1}, this region occurs for some solutions, which are, nonetheless, stable~\cite{oEMS1}.

%
\section{Results for specific examples with $f(\phi)=1$}\label{S7}
%
\par \par 
Let us now apply the foregoing formalism to particular cases. We shall focus on cases for which scalarisation is sourced by the non-minimal coupling to the $F_{\mu \nu}\tilde{F}^{\mu \nu}$ term, keeping $f(\phi)=1$. Cases for which $f(\phi)\neq 1$ and $h(\phi)=0$ have been considered more extensively in the literature, see $e.g.$~\cite{Gibbons:1987ps,Garfinkle:1990qj,SSChargedBH1,SSChargedBH2,EMSdyons}.

\subsection{A flat spacetime toy model}
\par Following~\cite{SSChargedBH1}, we start by presenting a flat spacetime toy model that illustrates the spontaneous scalarisation phenomenon with an axionic-type coupling and, moreover, shows that gravity is not fundamental for the phenomenon to occur. 

Consider the particular case where the background is the flat Minkowski spacetime, $i.e.$~\eqref{ma} with $N(r)=1$ and $\delta=0$. The corresponding Maxwell-scalar model allows a scalar-free configuration, which is simply the dyonic Coulomb configuration, given by
\begin{equation}
    \phi=0\ ,\, \qquad A=\frac{Q}{r} dt - P \cos{\theta} d\varphi \ .
\end{equation}
The scalar field equation of motion~\eqref{c4:eq:scalarfield} reduces to
\begin{equation}
    (r^2 \phi^\prime)^\prime - \frac{\dot h(\phi) P}{r^2}\left(Q+Ph(\phi)\right) = 0 \ .
\end{equation}
Introducing $\gamma(\phi)\equiv \left(Q+Ph(\phi)\right)^2$ and a new radial coordinate $x=1/r$, one arrives at
\begin{equation}
    \frac{d^2\phi}{dx^2}-\frac{1}{2} \frac{d\gamma(\phi)}{d\phi} = 0\ ,
\end{equation}
which takes the form of a one dimensional mechanical problem of a particle in a potential. The corresponding motion is described by the Lagrangian $L=T-U$, with $T=\left( d\phi/dx \right)^2$ and $U=-\gamma(\phi)$. We notice the existence of a first integral $E_0=T+U=\left( d\phi/dx \right)^2 - \gamma(\phi)$, from which analytical solutions for the radial profile of the scalar field may be obtained by solving the integral
\begin{equation}
    x=\int \frac{d\phi}{\sqrt{E_0+\gamma(\phi)}}=\int \frac{d\phi}{\sqrt{E_0+(Q+Ph(\phi))^2}} \ ,
    \label{c4:eq:toymodel}
\end{equation}
and then inverting the solution to obtain $\phi(x)$ and $\phi(r)$. 

As an explicit example, consider the coupling function
\begin{equation}
    h(\phi)=\frac{\sqrt{Q^2-\phi^2}-Q}{P} \ ,
\end{equation}
that respects all the previously discussed conditions for the occurrence of spontaneous scalarisation. For this particular choice of coupling function, which does not have any particular motivation rather than illustrating the phenomenon with a simple solution, it is possible to obtain a closed form spherically symmetric scalarised solution of the Maxwell-scalar system,
\begin{equation}
    \phi = \frac{\sqrt{E_0+Q^2} \tan\left(\frac{1}{r}\right)}{\sqrt{1+\tan\left(\frac{1}{r}\right)^2}},\, \qquad A= Q \, Ell\left(\frac{1}{r}, 1 + \frac{E_0^2}{Q^2}\right) dt - P \cos{\theta} d\varphi \ ,
    \label{c4:eq:toyscalarisedsol}
\end{equation}
where $Ell$ denotes the elliptic integral of second kind. This solution co-exists with the scalar-free dyonic Coulomb solution and one may inquire which one is preferred. To address this point, and since the Coulomb solution has infinite energy due to the singularity at the location of a point-charge, we consider a cut-off in the form of a conducting sphere located at $r=r_0$. The energy of any of the configurations can be computed as
\begin{equation}
    E=-2\pi \int_{r_0}^{\infty} \int_{0}^{\pi} T^t_t \, r^2 \sin{\theta} d\theta dr = 2\pi \int_{r_0}^{\infty} \int_{0}^{\pi} \rho \, r^2 \sin{\theta} d\theta dr \ .
\end{equation}
Denoting by $E^{\phi\neq0}$ and by $E^{\phi=0}$ the energy of the scalarised and scalar-free configurations respectively, one obtains
\begin{equation}
    E^{\phi\neq0}-E^{\phi=0}=4\pi(E_0^2+Q^2) \sin{\left( \frac{2}{r_0} \right)} \ ,
\end{equation}
hence, the scalarised solution if energetically favoured in a set of bands given by
\begin{equation}
    \frac{1}{r_0} \in \pi ] n+1 / 2, n+1[ \ ,
\end{equation}
where $n \in \mathbb{N}_{0}$, besides labelling the bands, counts the number of nodes exterior to $r_0$ of the scalar field profile. As advertised, this toy model shows spontaneous scalarisation through an axionic-type coupling: $(i)$ is not exclusive of gravitational theories; $(ii)$ may lead to favoured configurations.

%
\subsection{BHs with axionic hair ($h(\phi)=-\alpha\phi$)}
%
\par Depending on the choice of the coupling function $h(\phi)$, two sorts of models are possible: models in which the dyonic RN BH of electrovacuum is a solution, and models in which it is not. In~\cite{EMSdyons}, a classification of EMS models --  without an axionic coupling -- was performed based on this criterion. Here, we shall illustrate these two sorts of models with two examples of $h(\phi)$. In this subsection we choose $h(\phi)=-\alpha\phi$, which is the standard axionic coupling, for which the dyonic RN BH is not a solution and there are new BHs, all of which have axionic hair. In this case the scalar field $\phi$ is dubbed \textit{axion}.

The BHs of the model with $h(\phi)=-\alpha\phi$  were first constructed in~\cite{Weinberg}; our analysis will complement the study therein. Since all BH solutions are scalarised, no bifurcation points from the scalar-free RN BH exist. We will study the domain of existence of such BHs with axionic hair, which has not been presented before. For particular solutions, the radial profiles will be compared with the results in~\cite{Weinberg} obtaining agreement. We shall also obtain, for these BHs with axionic hair, the effective potential for spherical perturbations.

\subsubsection{Radial profiles}
\par
\begin{figure}[h!]
    \centering
    \includegraphics[width=.43\textwidth]{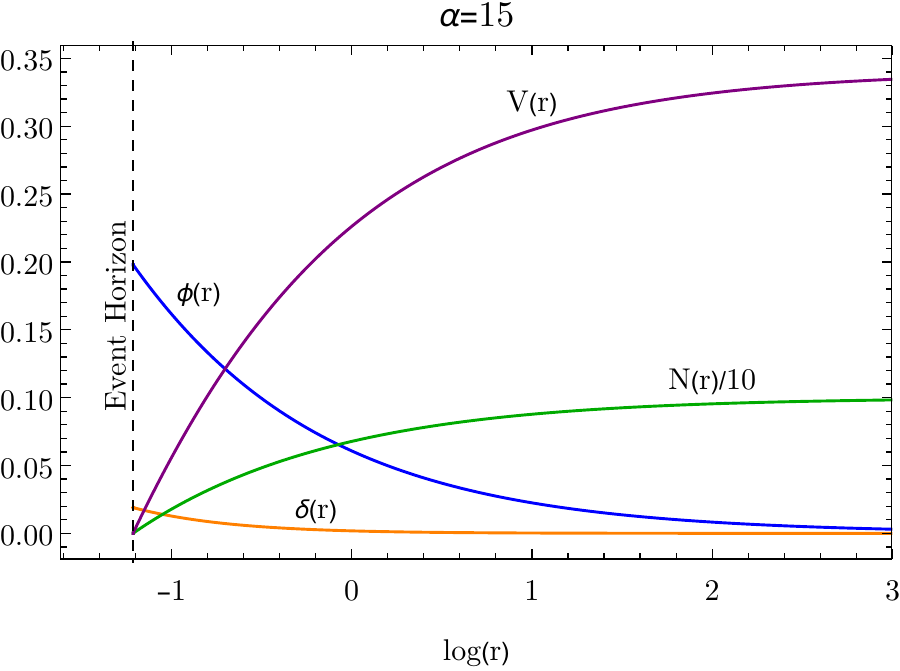}\hfill
    \includegraphics[width=.43\textwidth]{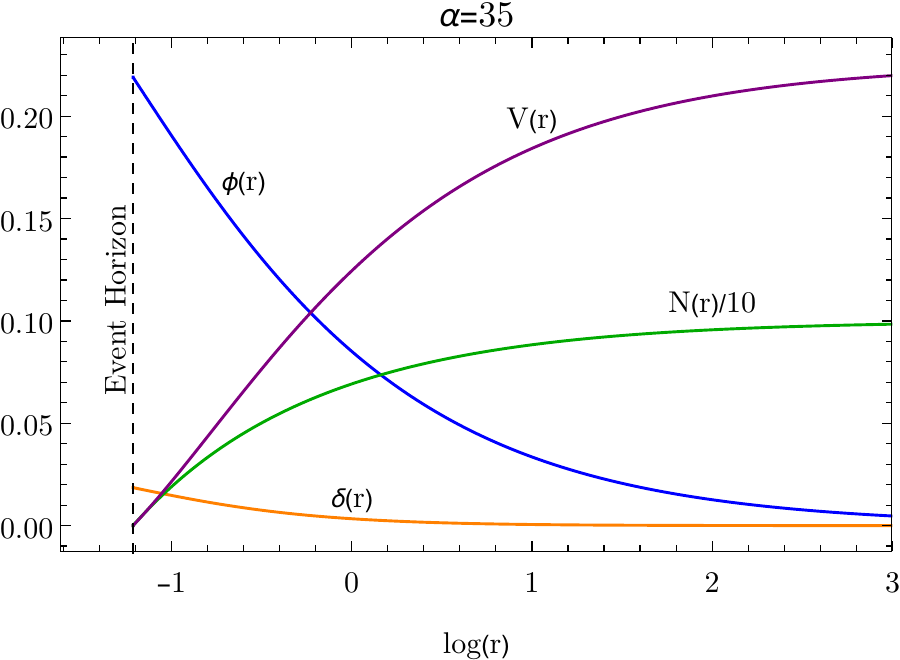}
    \caption{Radial functions for BHs with axionic hair. $(Q,P,r_H)=(0.12,0.012,0.2974)$ and two values of $\alpha$.}
    \label{c4:fig:WeinProfiles}
\end{figure}

 Some typical solutions of the various functions that define scalarised BHs obtained from numerical integration can be found in Fig. \ref{c4:fig:WeinProfiles} for two illustrative values of the coupling constant $\alpha=15,35$, while keeping $Q$, $P$ and $r_H$ constant. Some characteristic quantities can be found in Table \ref{c4:tab:WeinRadialProfiles}.

\begin{table}[H]
    \centering
    \small
    \begin{tabular}{c|c|c|c|c|c|c|c|c|c|c}
        $\alpha$ & $P$ & $\beta$ & $q$ & $M$ & $Q_s$ & $\Phi_e$ & $\Phi_m$ & $U$ & $a_H$ & $t_H$ \\ \hline
        15 & 0.006 & 0.05 & 0.699124 & 0.171858 & 0.034413 & 0.38440 & 0.0201273 & -0.00204597 & 0.748456 & 0.99244 \\
        15 & 0.012 & 0.1 & 0.711658 & 0.169461 & 0.061265 & 0.34058 & 0.0400918 & -0.00579648 & 0.769778 & 1.02653 \\
        35 & 0.006 & 0.05 & 0.713878 & 0.168306 & 0.068077 & 0.32431 & 0.0200267 & -0.00685364 & 0.780381 & 1.03639 \\
        35 & 0.012 & 0.1 & 0.742068 & 0.162517 & 0.094770 & 0.22572 & 0.0400274 & -0.00946308 & 0.836965 & 1.06170
    \end{tabular}
    \caption{Properties of BHs with axionic hair ($r_H=0.29736$).}
    \label{c4:tab:WeinRadialProfiles}
\end{table}
\par All BHs with axionic hair we have constructed have an axion profile which is nodeless. The axion is a monotonically decreasing function of the radius; hence $\phi_0$ is always the maximum value of the axion field. Data reveal that (as expected) for larger couplings $\alpha$, there is a higher degree of scalarisation, as seen from the value of $\phi_0$ in Fig. \ref{c4:fig:WeinProfiles} and in the values of the scalar charge (Table \ref{c4:tab:WeinRadialProfiles}). 

\begin{figure}[ht!]
\centering
\includegraphics[width=.43\textwidth]{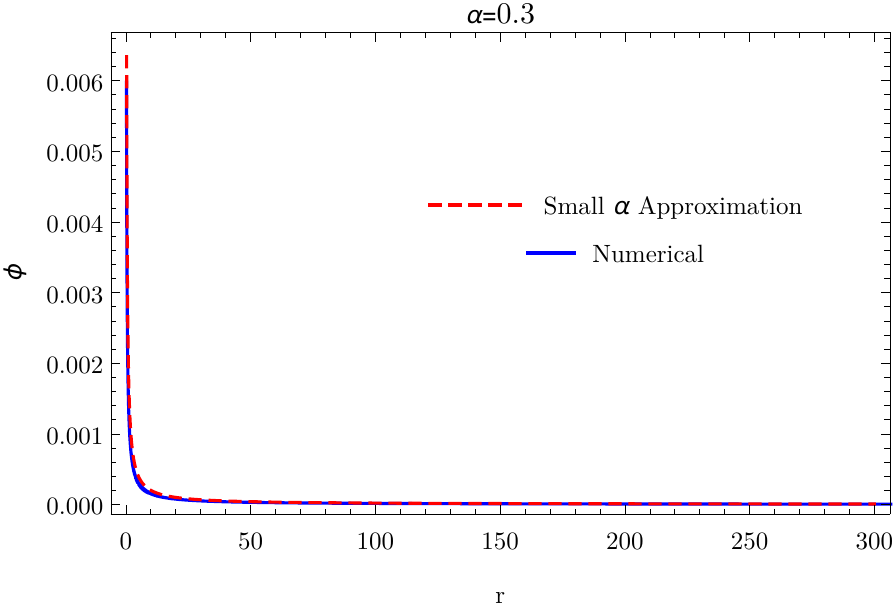}\hfill
\includegraphics[width=.43\textwidth]{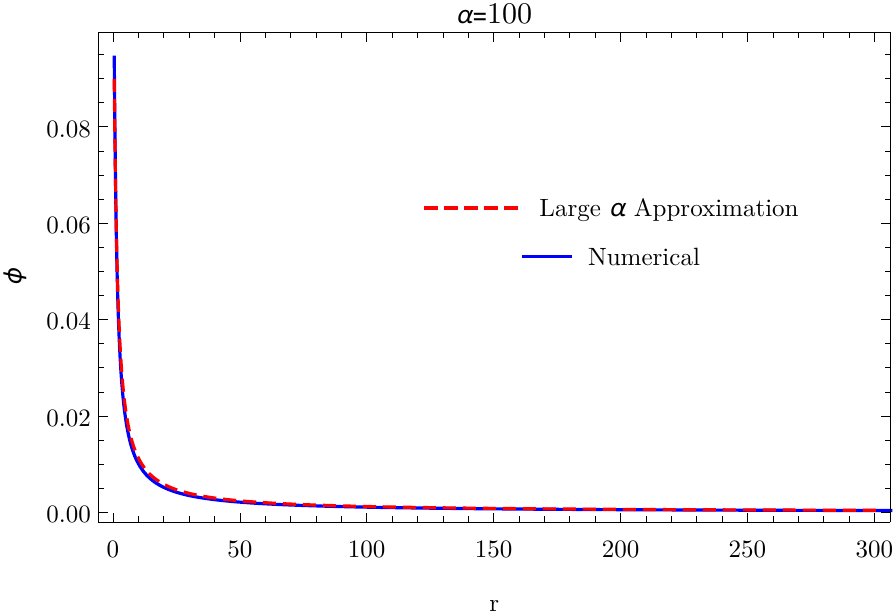}\vfill
\includegraphics[width=.43\textwidth]{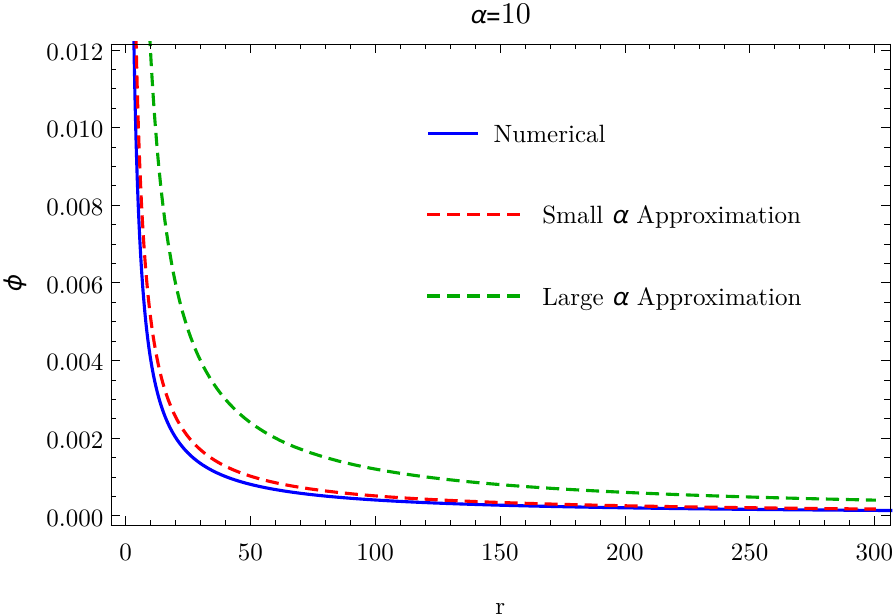}\hfill
\includegraphics[width=.43\textwidth]{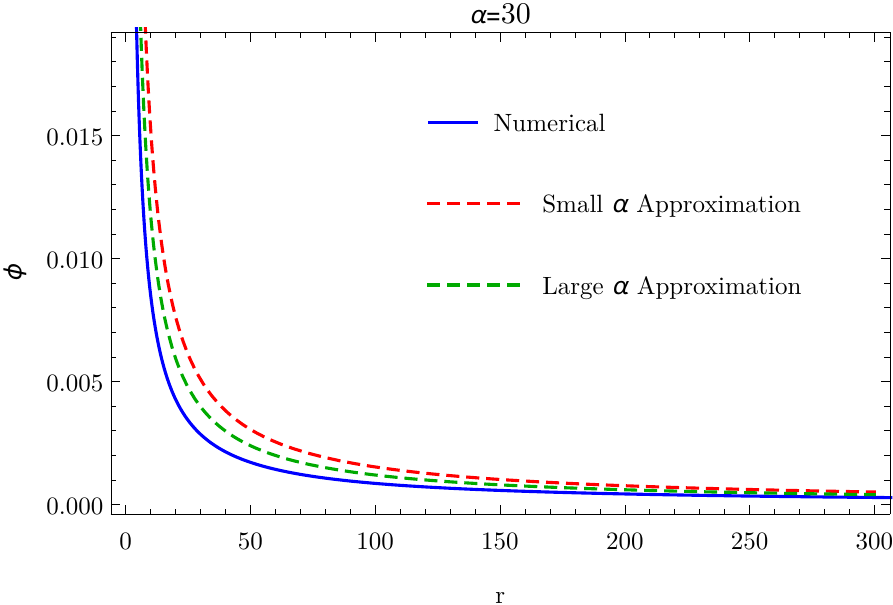}
\caption{Analytical approximations $vs.$ numerical results. (Top left) $\alpha=0.3$, $(Q,P,r_H)=(0.12,0.012,0.28247)$, approximation for small $\alpha$. (Top right) $\alpha=100$, $(Q,P,r_H)=(0.12,0.012,0.52966)$, approximation for large $\alpha$. (Bottom left) $\alpha=10$, $(Q,P,r_H)=(0.12,0.012,0.33453)$. (Bottom right) $\alpha=30$, $(Q,P,r_H)=(0.12,0.012,0.33453)$. The approximations hold well in their regime of validity. The intermediate cases exhibit a visible mismatch between the numerical solution and the analytical approximations.}
\label{c4:fig:WeinApprox}
\end{figure}

In \cite{Weinberg} two approximations for the axion radial profile were derived from eq. \eqref{c4:eq:scalarfield} as follows
\begin{equation}
    \phi(r) \approx \alpha \frac{QP}{r_+ r_-} \ln{\left(\frac{r}{r-r_-}\right)} + \mathcal{O}(\alpha^2), \qquad  \text{for small $\alpha$} \ ,
\end{equation}
\begin{equation}
    \phi(r) \approx \frac{Q}{\alpha P} \left(1-\exp{\left(-\frac{\alpha P}{r}\right)}\right), \qquad \text{for large $\alpha$} \ ,
\end{equation}
where $r_{\pm}=M\pm \sqrt{M^2-Q^2-P^2}$ denote the radial coordinate of the horizons of the scalar-free RN BH. A comparison between these analytical approximations and the numerical results can be seen in Fig. \ref{c4:fig:WeinApprox}.

\par
\begin{figure}[ht!]
\centering
\includegraphics[width=.45\textwidth]{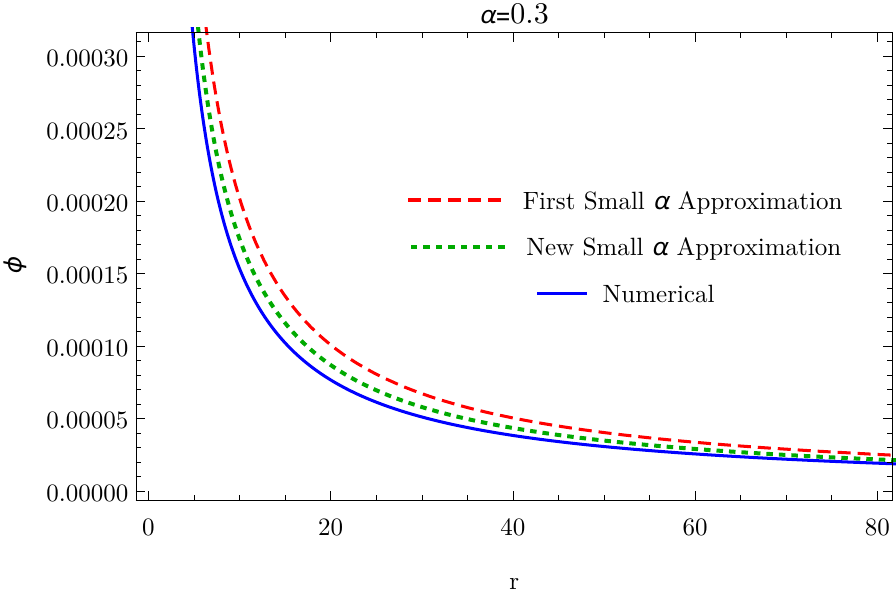}\hfill
\includegraphics[width=.45\textwidth]{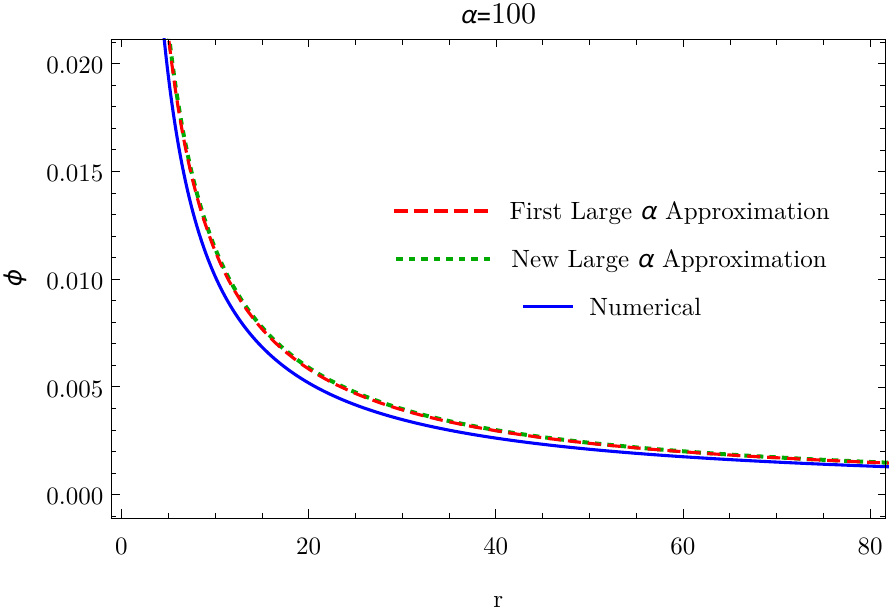}
\caption{Zoom in on the axion radial profile, in order to compare the new and original analytical approximations, together with the numerical results. (Left) $\alpha=0.3$, $(Q,P,r_H)=(0.12,0.012,0.28247)$. (Right) $\alpha=100$, $(Q,P,r_H)=(0.12,0.012,0.52966)$.}
\label{c4:fig:WeinApproxNew}
\end{figure}

 We have introduced two improved analytical approximations for large and small couplings (in the same spirit as in \cite{Weinberg}) that besides taking into account the value $\phi_0$ for the axion at the event horizon, in the small coupling limit does not neglect $\alpha^2$ terms. These approximations are
\begin{equation}
    \phi(r) \approx \frac{1}{\alpha \beta} + A\, LP{_v}\left(- \frac{(Q^2+P^2)(r-2r_H)+r_H^2 r}{r(Q^2+P^2-r_H^2)}\right)+B\, LQ{_v}\left(- \frac{(Q^2+P^2)(r-2r_H)+r_H^2 r}{r(Q^2+P^2-r_H^2)}\right), \, \text{for small $\alpha$} \ ,
\end{equation}
where $LP$ and $LQ$ are Legendre functions of first and second kind, respectively, and $v=\frac{1}{2} \left(\sqrt{\frac{4 \alpha^2 \beta^2}{1+\beta^2}+1}-1\right)$. $A$ and $B$ are integration constants chosen to guarantee an asymptotically vanishing solution and that the horizon value of the axion is $\phi_0$. The opposite approximation is
\begin{equation}
    \phi(r) \approx \frac{1}{\alpha \beta}\left(1+\frac{\left(\phi_0 \alpha \beta - 1\right) \sinh{\left(\frac{\alpha P}{r} \right) + \sinh{\left(\alpha P \left(\frac{1}{r}-\frac{1}{r_H} \right)\right)}} }{\sinh{\left(\frac{\alpha P}{r_H} \right)} }\right), \qquad \text{for large $\alpha$} \ .
\end{equation}
These new approximations are compared with the previous ones and the numerics in Fig.~\ref{c4:fig:WeinApproxNew}. Data reveals that, indeed, the new approximations hold better than the previous ones, with the differences being more relevant in the small coupling case. Although less obvious from Fig. \ref{c4:fig:WeinApproxNew}, numerics indicate that the new large coupling approximation also gives an overall better approximation. As code tests on the numerical solutions, we report differences of $10^{-9}$ for the Virial relation, and $10^{-8}$ for the Smarr law.

\subsubsection{Domain of Existence}
\par
\begin{figure}[h!]
\centering
\includegraphics[width=.45\textwidth]{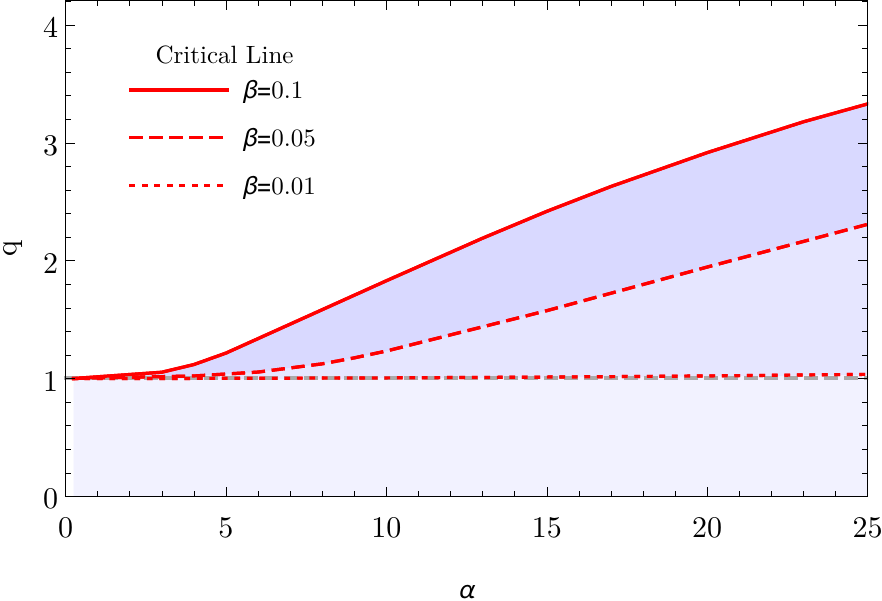} \hfill
\includegraphics[width=.45\textwidth]{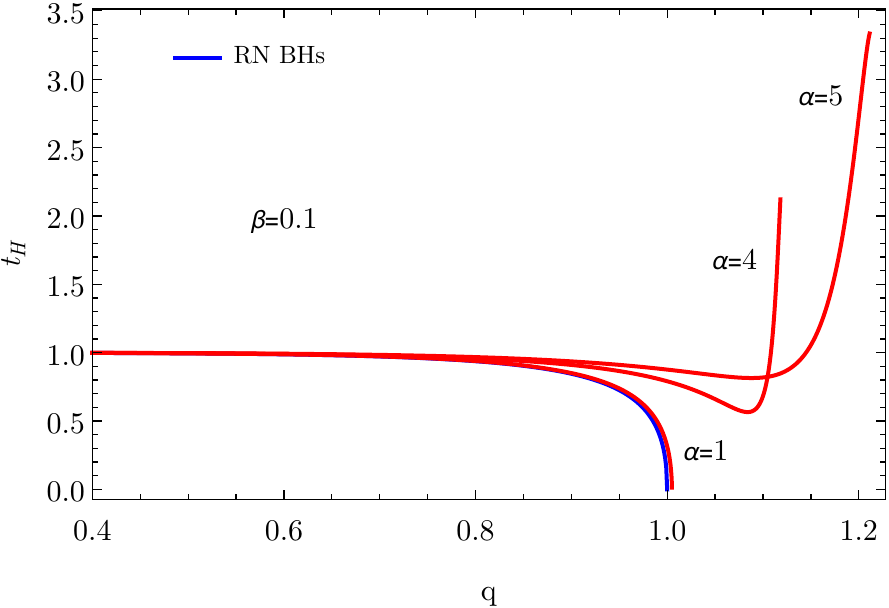}
\caption{(Left) Domain of existence of BHs with axionic hair in the $(\alpha,q)$ plane (blue shaded regions plus the $q<1$ region). For each $\beta$ value, the domain is bounded by a critical line, at which solutions become singular. (Right) Reduced temperature $t_H$ vs $q$ for a sample of $\alpha$ values.}
\label{c4:fig:WeinDomain}
\end{figure}

 The domain of existence, in the $(\alpha,q)$ plane, for BHs with axionic hair is presented in Fig. \ref{c4:fig:WeinDomain} (left panel). It is delimited by a set of critical solutions that we call the \textit{critical line}. At the critical line numerics suggest a divergence of the Kretschmann scalar and of the horizon temperature, together with a vanishing of the horizon area. Simultaneously, $M$ and $Q_s$ remain finite and non-zero. However, for sufficiently small values of the coupling, for which the axionic hair is almost negligible, some physical quantities  behave similarly to those for an extremal RN BH, $e.g.$ the event horizon temperature (\textit{cf.} Fig. \ref{c4:fig:WeinDomain}, right panel). It is interesting to notice that, along $\alpha=constant$ branches, $q$ increases beyond unity: therefore, BHs with axionic hair can be overcharged. 

For a given $\alpha$, as $q$ increases, so does the axion's initial amplitude $\phi_0$. Thus, the global maximum of the axion, which always occurs at the horizon, increases, for fixed $\alpha$, with $q$. This means one may take $\phi_0$ as a measure of $q$ and vice-versa. As another feature, fixing $Q$, there is a wider domain of existence for larger $P$. This behaviour contrasts with the one observed for the dyonic scalarised BHs in \cite{EMSdyons} (which has $h(\phi)=0$ and $P\neq 0$). This wider domain of existence, for larger $\beta$, may be associated to the fact that the coupling $h(\phi) F_{\mu \nu} \Tilde{F}^{\mu \nu}$ is directly proportional to $P$; thus, for larger values of $\beta$ the axion couples more strongly to the source term.

\subsubsection{Effective potential for spherical perturbations}
\par
\begin{figure}[ht!]
\centering
\includegraphics[width=.4\textwidth]{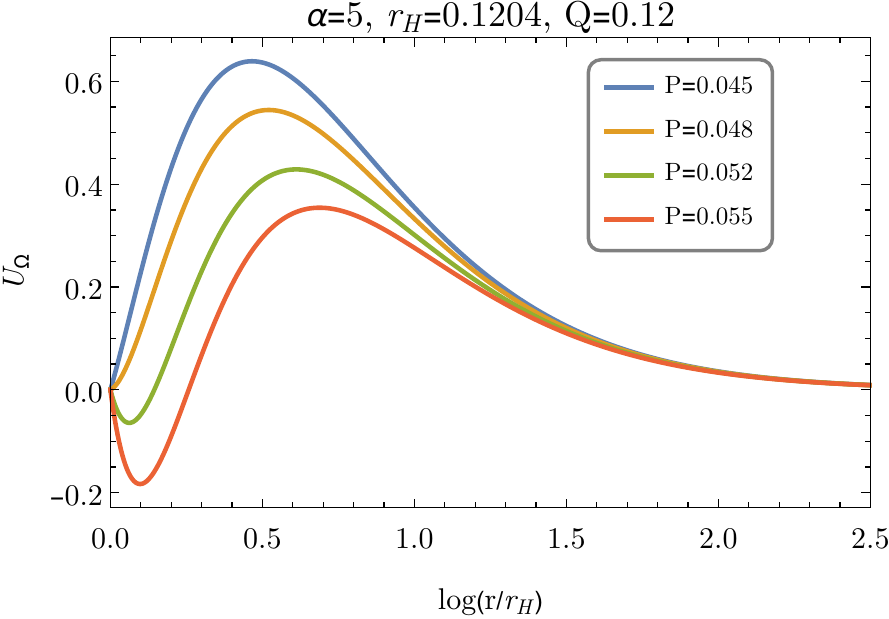}\hfill
\includegraphics[width=.4\textwidth]{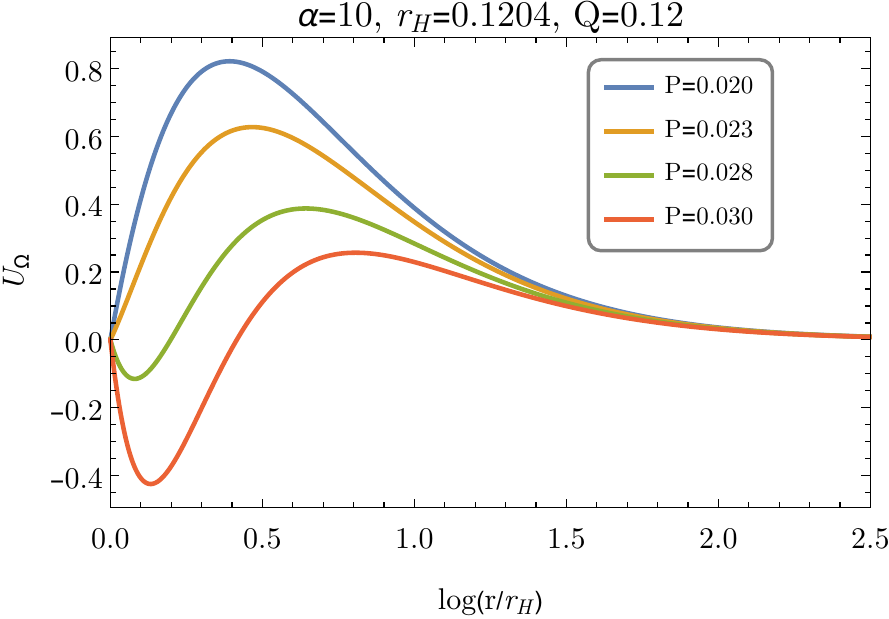}
\caption{Effective potential for spherical perturbations, $U_\Omega$,  for BHs with axionic hair. The sample of solutions have $r_H=0.1204$, $Q=0.12$ and $\alpha=5$ (left panel) or $\alpha=10$ (right panel). The potential is always positive until a critical $\beta$ value is reached, beyond which a negative region appears.}
\label{c4:fig:WeinPert}
\end{figure}

 The effective potential for spherical perturbations $U_\Omega$, for a sample of BHs with axionic hair is plotted in Fig. \ref{c4:fig:WeinPert}. It is not positive definite in all cases but it is regular in the entire range $-\infty<x<\infty$. For other values of the coupling $\alpha$, the potential always behaves in a similar way: for sufficiently small values of $\beta$, the potential is always positive (and hence, free of instabilities), until $\beta$ reaches a value at which the potential starts to have a negative region. The potential vanishes at the horizon and at infinity.

\subsection{Spontaneous scalarisation with an axionic-type coupling ($h(\phi)=-\alpha\phi^2$)}
\par We now turn to the second sort of models, for which the dyonic RN BH of electrovacuum is a solution, but new, dubbed scalarised BHs, are also possible. In these models the phenomenon of spontaneous scalarisation may occur. To illustrate it we consider the coupling function $h(\phi)=-\alpha \phi^2$ (and $f(\phi)=1$ as before), hereby dubbed ``power-law coupling". The coupling constant $\alpha$ is a dimensionless positive constant. This coupling function obeys all the necessary conditions for the occurrence of spontaneous scalarisation and it is the natural generalisation of the axionic case. Again, we shall present radial profiles and the domain of existence of the scalarised BHs. Moreover, since the RN BH is a solution of the model (the scalar-free solution), an entropic analysis will be performed which shows the scalarised solutions are entropically favoured. Then, a fully non-linear dynamical analysis will show that the scalar-free solutions indeed dynamically evolve to form scalarised BHs.

\subsubsection{Radial profiles}
\par
\begin{figure}[ht!]
    \centering
    \includegraphics[width=.4\textwidth]{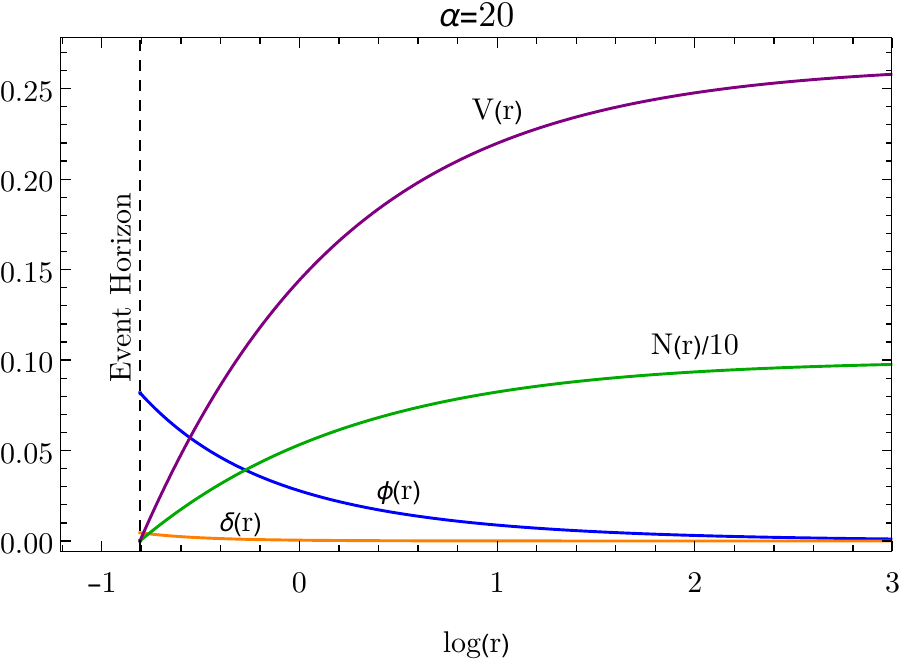}\hfill
    \includegraphics[width=.4\textwidth]{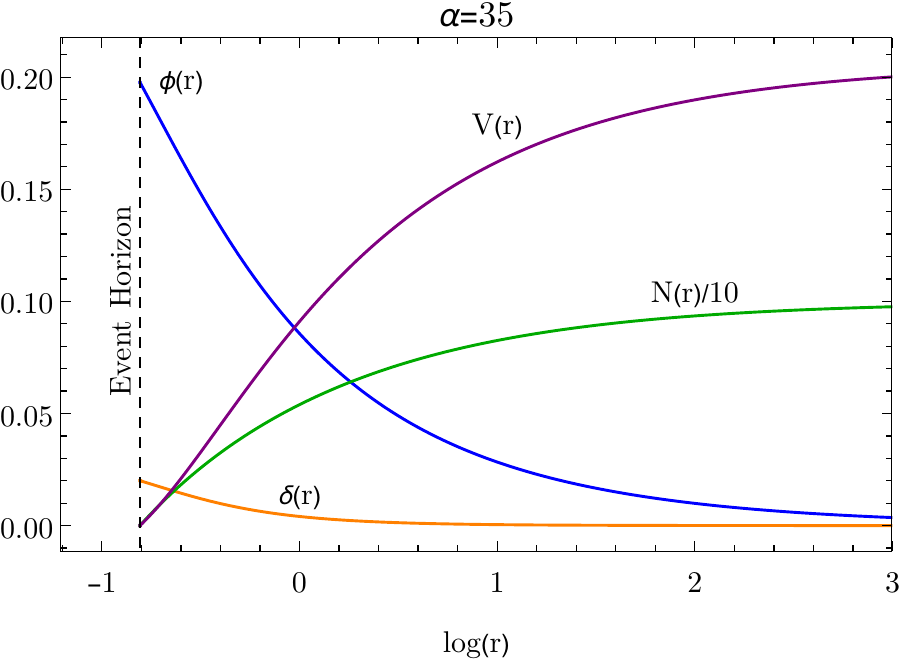}\vfill
    \includegraphics[width=.4\textwidth]{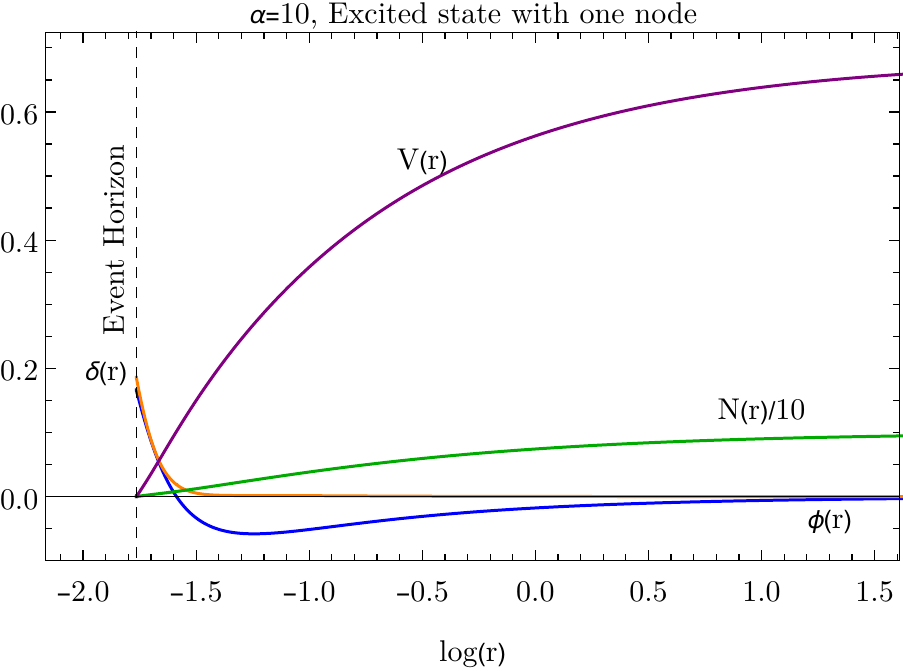}\hfill
    \includegraphics[width=.4\textwidth]{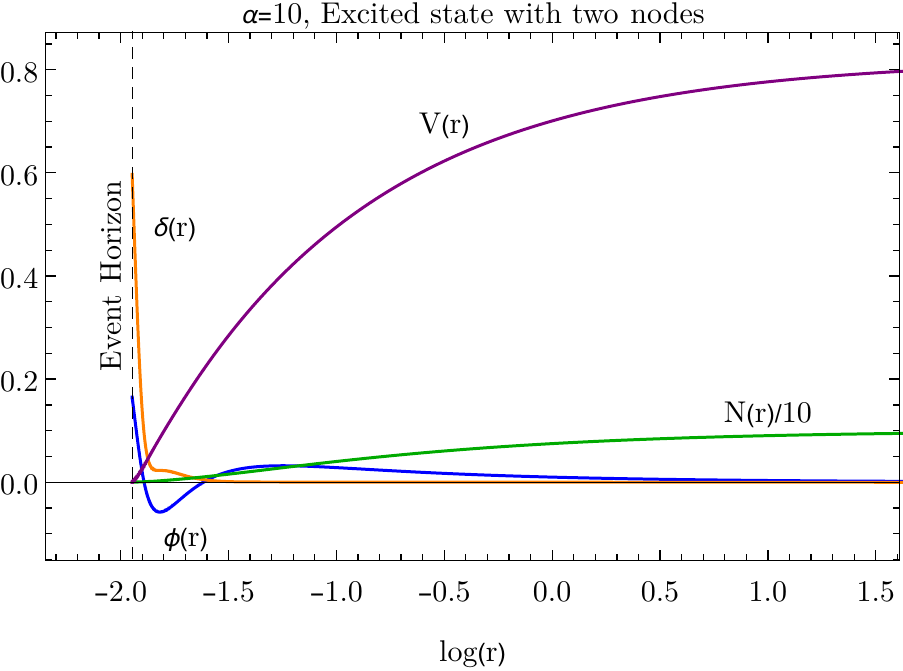}
    \caption{(Top) Radial functions for scalarised BHs in the power-law coupling model, with $\alpha=20$ (left) and $\alpha=35$ (right), with $(Q,P,r_H)=(0.12,0.06,0.44604)$. (Bottom left) Excited solution with $n=1$, $\alpha=10$ and $(Q,P,r_H)=(0.12,0.06,0.171279)$. (Bottom right) Excited solution with $n=2$, $\alpha=10$ and $(Q,P,r_H)=(0.12,0.06,0.142733)$.}
    \label{c4:fig:PowerProfiles}
\end{figure}

 Some typical profiles for the various functions that define scalarised BHs, in the power-law coupling model, obtained from numerical integration, can be found in Fig. \ref{c4:fig:PowerProfiles}, for two illustrative values of the coupling constant, $\alpha=20,35$, while keeping $Q$, $P$ and $r_H$ constant (some characteristic quantities can be found in Table \ref{c4:tab:PowerRadialProfiles}). Two excited solutions (with $n=1$ and $n=2$) are also presented.%
\begin{table}[ht!]
    \centering
    \footnotesize
    \begin{tabular}{c|c|c|c|c|c|c|c|c|c|c}
        $\alpha$ & $P$ & $\beta$ & $q$ & $M$ & $Q_s$ & $\Phi_e$ & $\Phi_m$ & $U$ & $a_H$ & $t_H$ \\ \hline
        20 & 0.06 & 0.5 & 0.5516939 & 0.2431857 & 0.02174461 & 0.263942 & 0.134363 & -0.000552335 & 0.84103 & 0.99744 \\
        20 & 0.072 & 0.6 & 0.5719615 & 0.244672 & 0.04789236 & 0.242883 & 0.160661 & -0.00242039 & 0.830845 & 1.01522 \\
        35 & 0.06 & 0.5 & 0.5553367 & 0.2415905 & 0.07080436 & 0.206099 & 0.133572 & -0.00422197 & 0.852173 & 1.03492 \\
        35 & 0.072 & 0.6 & 0.5769590 & 0.2425525 & 0.07730736 & 0.190191 & 0.160293 & -0.00452185 & 0.845427 & 1.0373
    \end{tabular}
    \caption{Properties of illustrative scalarised BHs in the power-law coupling model.} 
    \label{c4:tab:PowerRadialProfiles}
\end{table}

\par Some qualitative features observed for the BHs with axionic hair remain here. For the nodeless solutions $\phi_0$ is the maximum of the scalar field radial profile. Larger couplings imply stronger scalarisation and smaller electrostatic potentials. The excited states of Fig. \ref{c4:fig:PowerProfiles} (bottom panel) have larger $\delta_0$ and electrostatic potentials when compared to the fundamental solutions. 

\begin{figure}[ht!]
\centering
\includegraphics[width=.5\textwidth]{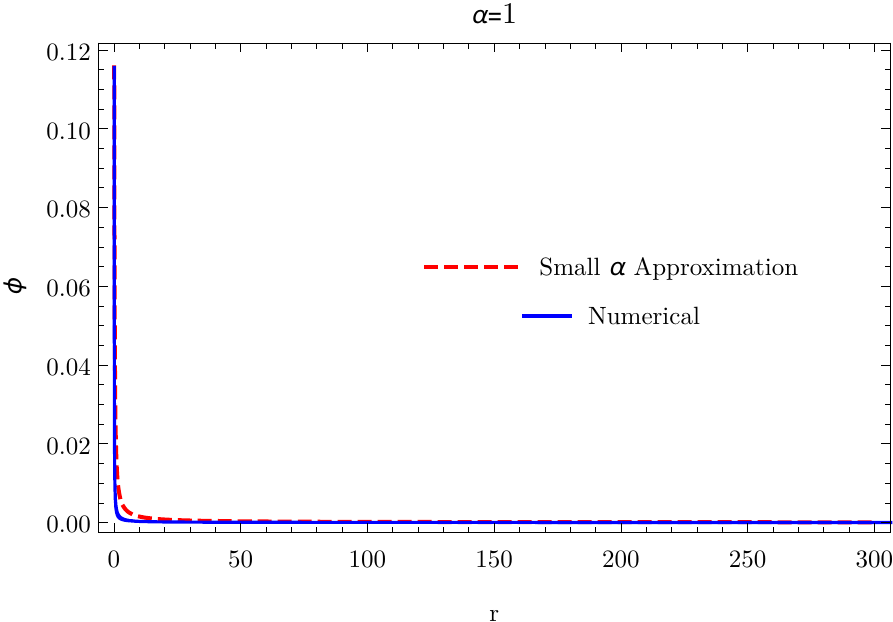}\hfill
\includegraphics[width=.5\textwidth]{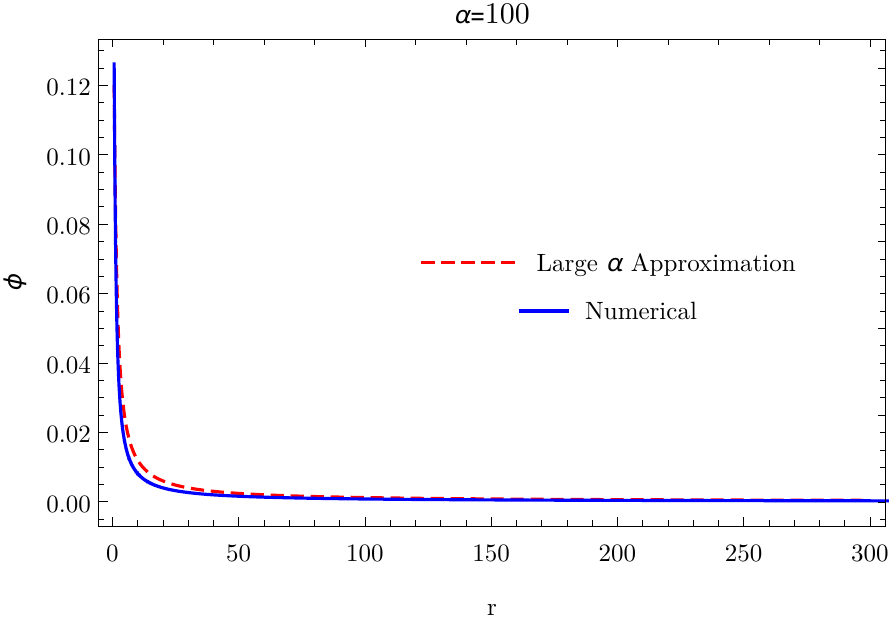}
\caption{Analytical approximations $vs.$ numerical results. (Left) $\alpha=1$, $(Q,P,r_H)=(0.12,0.06,0.14273)$, approximation for small $\alpha$. (Right) $\alpha=100$, $(Q,P,r_H)=(0.12,0.06,0.68393)$, approximation for large $\alpha$.}
\label{c4:fig:PowerApprox}
\end{figure}

From eq. \eqref{c4:eq:scalarfield}, it is also possible to derive two approximations, for small and large couplings, in a similar fashion as it was done for the case of the axionic coupling:
\begin{equation}
    \phi(r) \approx A\, LP{_u}\left(- \frac{(Q^2+P^2)(r-2r_H)+r_H^2 r}{r(Q^2+P^2-r_H^2)}\right)+B\, LQ{_u}\left(- \frac{(Q^2+P^2)(r-2r_H)+r_H^2 r}{r(Q^2+P^2-r_H^2)}\right) + \mathcal{O}(\alpha^2), \, \text{for small $\alpha$},
\end{equation}
where $LP_u$ and $LQ_u$ are Legendre functions of first and second kind, respectively, and $u=\frac{1}{2} \left(\sqrt{1-\frac{8 \alpha \beta}{1+\beta^2}}-1\right)$. Again, $A$ and $B$ are integration constants chosen to guarantee the correct asymptotic behaviours. For large couplings we have
\begin{equation}
        \phi(r) \approx \frac{1}{\sqrt{\alpha \beta}} \tanh \left(\frac{\sqrt{\alpha P Q}}{r}\right), \, \text{for large $\alpha$} \ ,
\end{equation}
where the integration constants were chosen such that the solution is regular and asymptotically vanishing. A comparison between these analytical approximations and the numerical results can be found in Fig. \ref{c4:fig:PowerApprox}. The approximations are qualitatively worse than in the  axionic case. Likely, this is due to the non-linearities induced by the quadratic scalar field terms. The small coupling approximation neglects $\alpha^2$ terms and the large coupling one does not possess the benefit of imposing the scalar field value at the event horizon, since one of the integration constants was used to impose regularity of the approximation. 
We remark that, once again, tests to the code reveal relative differences $10^{-9}$ for the Virial relation, $10^{-7}$ for the Smarr law.

\subsubsection{Domain of existence}
\par In the case studied in the previous section, dyonic BHs with axionic hair, the domain of existence was, generically, bounded by a critical line: a set of singular solutions in which the horizon shrinks to a point. This is also the case for the purely electrically charged scalarised BHs studied in \cite{SSChargedBH1,SSChargedBH2}. A different behaviour was observed for the dyonic BHs in~\cite{EMSdyons}. The domain of existence is then bounded by an extremal line: a set of solutions for which the horizon temperature tends to zero, while the Kretschmann scalar and the horizon area remain finite and non-zero. We will now see that for the dyonic scalarised BHs with the power-law coupling herein the latter behaviour also occurs.

\begin{figure}[ht!]
\centering
\includegraphics[width=.5\textwidth]{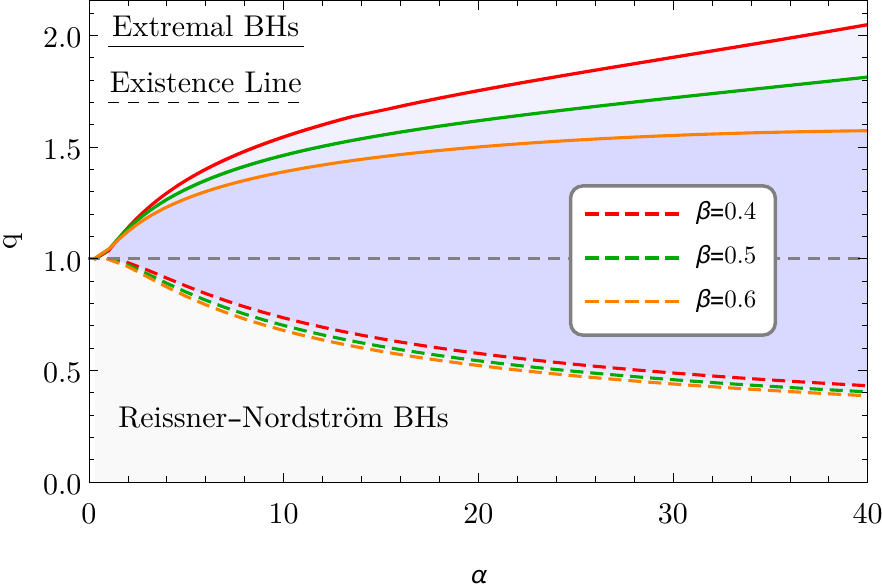}
\caption{Domain of existence of scalarised BHs in the power-law model, in the $(\alpha,q)$ plane. It is bounded by an extremal (solid) line, composed of extremal BHs, and by an existence (dashed) line, where the scalarised solutions bifurcate from the dyonic RN BH of electrovacuum.}
\label{c4:fig:PowerDomain}
\end{figure}

 In order to obtain the extremal BH solutions, a different near-horizon expansion which accounts for a degenerate horizon must be used (see \textit{e.g.} \cite{EMSdyons,ExtExpansionDilaton1,ExtExpansionDilaton2}). A double zero for $N(r)$ at the horizon is accommodated with the following expansion:
\begin{equation}
    \begin{array}{c}{N(r)=N_{2}\left(r-r_{H}\right)^{2}+\ldots \ ,} \\ {\phi(r)=\phi_{0}+\phi_{c}\left(r-r_{H}\right)^{k}+\ldots \ .}\end{array}
\end{equation}
Again the field equations related these parameters. We take $\phi_0$ and $r_H$ as the independent parameters. Then:
\begin{equation}
    P=r_H \, , \qquad Q=\alpha \phi_0^2 r_H \, , \qquad N_{2}=\frac{1}{r_{H}^{2}} \, , \qquad k=\frac{1}{2}\left(-1+\sqrt{1+16 \phi_{0}^{2} \alpha^{2}}\right) \ .
\end{equation}
A non-integer $k$ would imply that the derivative of all functions, at a sufficiently high order, diverge as $r\to r_H$. Then, suitable order derivatives of the Riemann tensor would diverge at the horizon. Here we focus on smooth solutions, requiring
\begin{equation}
    \alpha = \frac{\sqrt{p(p+1)}}{2\phi_0} \ , \qquad \, p \in \mathbb{N}_0 \ .
\end{equation}
In this way we obtained directly extremal solutions, which make up a boundary of the domain of existence. This domain is presented in Fig. \ref{c4:fig:PowerDomain}, in the $(\alpha,q)$ plane.

\par
\begin{figure}[ht!]
\centering
\includegraphics[width=.45\textwidth]{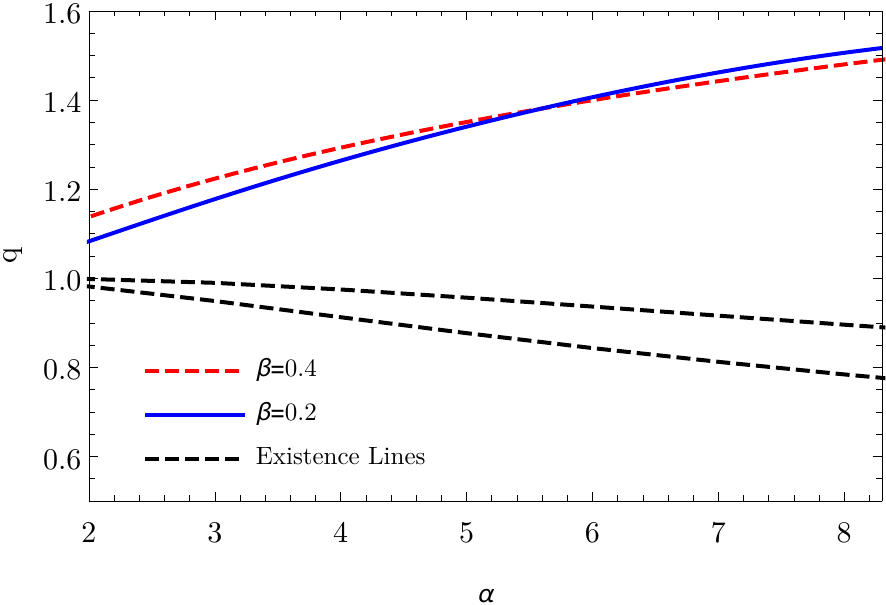}\hfill
\includegraphics[width=.45\textwidth]{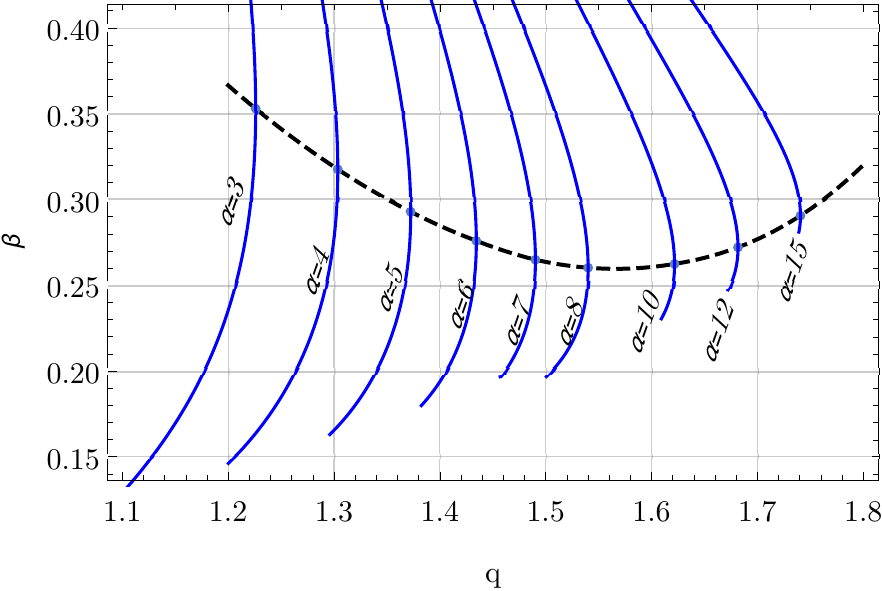}
\caption{(Left panel) Zoom in on the domain of existence for $\beta = 0.2,0.4$. There is a region wherein larger $\beta$ allows more overcharging. (Right panel) $(\beta,q)$ plane. The solid lines correspond to the extremal solution for each $\beta$ value, for a constant $\alpha$. The black curve interpolates the optimum value of $\beta$ for overcharging.}
\label{c4:fig:PowerDomainZoom}
\end{figure}

 The domain of existence reveals a region of non-uniqueness where, for the same charge to mass ratio $q<1$, RN BHs and scalarised BHs co-exist. Unlike in the domain of existence of the axionic case, for larger $\beta$ there is a narrower domain of existence. This trend, however, is not universal. For small enough $\alpha$, the domain of existence can be wider for larger $\beta$ - Fig. \ref{c4:fig:PowerDomainZoom} (left panel). 
Fig. \ref{c4:fig:PowerDomainZoom} (right panel) shows, in fact, a more subtle behaviour. The solid blue lines are extremal solution for $\alpha=$constant. One can observe that there is an optimal value of $\beta$ for overcharging. As $\alpha$ increases, however, this optimal value tends to the minimum value of $\beta$.

\subsubsection{Effective potential for spherical perturbations}
\par
\begin{figure}[ht!]
\centering
\includegraphics[width=.45\textwidth]{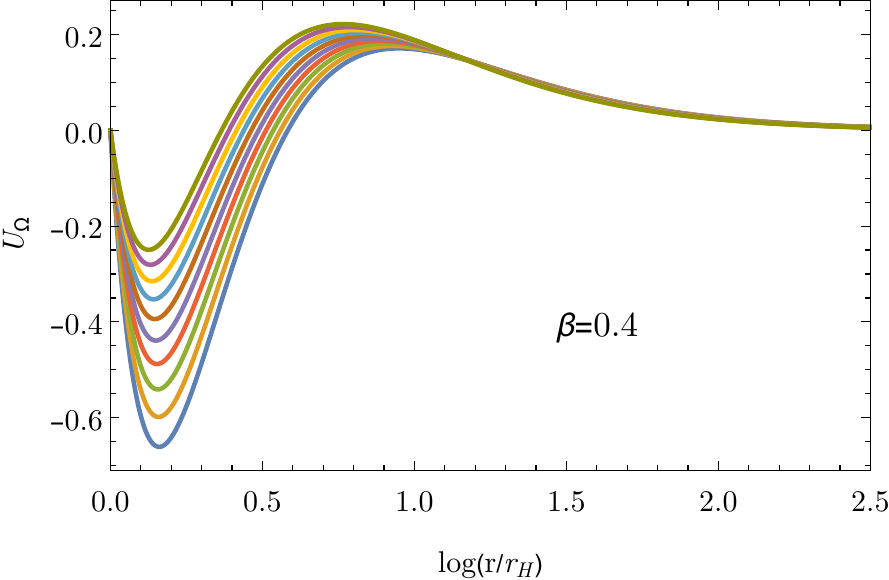}\hfill
\includegraphics[width=.45\textwidth]{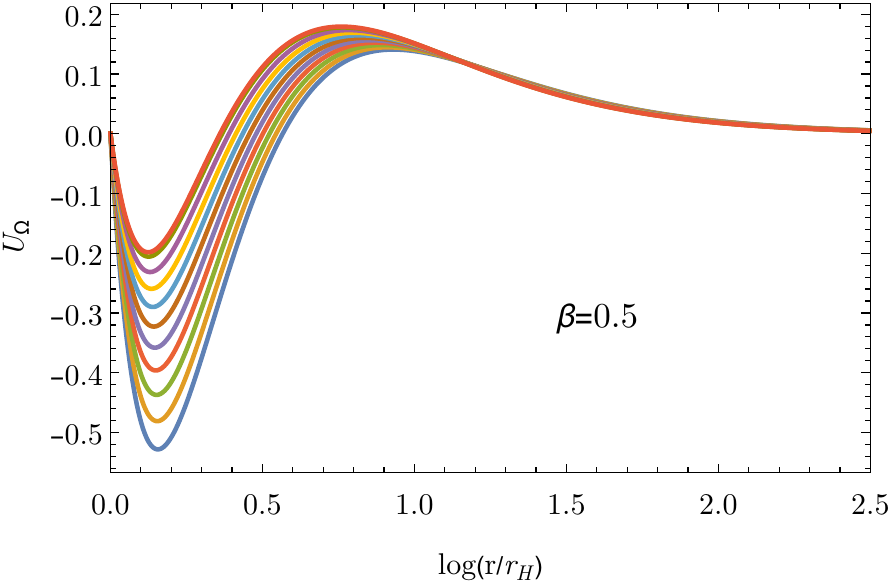}
\caption{Sequence of effective potentials, $U_\Omega$, for $\alpha=10$ and two values of $\beta$. There is always a region where the potential is negative. The deepest potentials occur for the largest $q$. The bottom/top curves occur for $q=0.785/0.736$ (left panel) or $q=0.746/0.701$ (right panel).}
\label{c4:fig:perturbative}
\end{figure}

 The effective potential, $U_\Omega$, for the scalarised BHs in the power-law coupling model is plotted in Fig. \ref{c4:fig:perturbative} for some illustrative examples. Again, it vanishes at the BH event horizon  and at infinity, but it is not positive definite. For other values of the coupling and $\beta$ ratios the potential always has a similar form. The existence of a region of negative potential does not imply instability. Thus, conclusions about linear stability require a study of quasi-normal modes, similarly to what was done in \cite{oEMS1} for the purely electrical case. Below, however, we shall present evidence, using fully non-linear numerical simulations that the scalarised solutions are not only stable, but in fact form dynamically.   

\subsubsection{Entropic and dynamical preference}
\par
\begin{figure}[ht!]
\centering
\includegraphics[width=.45\textwidth]{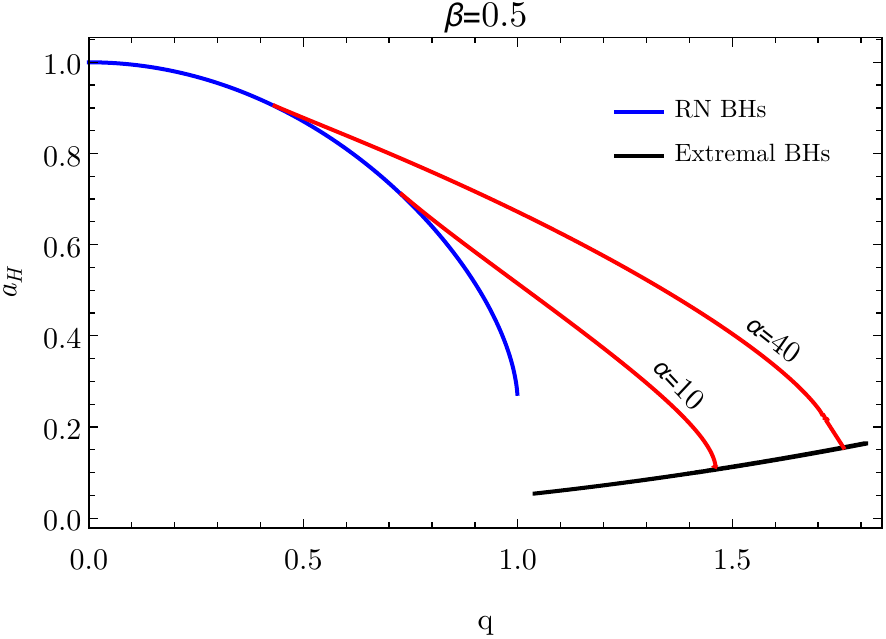}\hfill
\includegraphics[width=.45\textwidth]{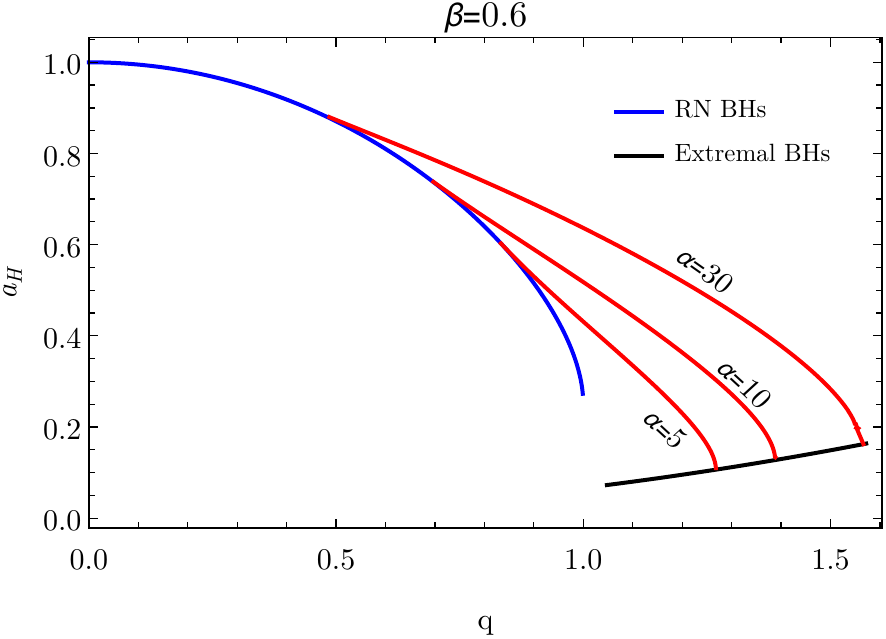}
\caption{$a_H$ $vs.$ $q$ for $\beta=0.5$ (left) and $\beta=0.6$ (right). The blue lines represent scalar-free RN BHs, while the red lines are sequences of (numerical data points representing) scalarised BHs for a given $\alpha$. The black line represents the entropy of the extremal BH solutions for the different $\alpha$ values.}
\label{c4:fig:PowerEntropy}
\end{figure}

 Since the model under consideration is General Relativity coupled to some matter, the Bekenstein-Hawking BH entropy formula holds. Thus, the entropy analysis reduces to the analysis of the horizon area. It is convenient to use the reduced event horizon area $a_H$. Then, in the region where the RN BHs and scalarised BHs co-exist - the non-uniqueness region -, for the same $q$, the scalarised solutions are always entropically preferred as seen in Fig. \ref{c4:fig:PowerEntropy}.

\par The entropic preference, together with the instability of the scalar-free RN BHs against scalaristation, suggest that the latter evolve towards the former, when perturbed, at least if the evolutions are approximately conservative. We have performed fully non-linear numerical evolutions to test this scenario, following our previous work~\cite{SSChargedBH1,SSChargedBH2}. The initial data is a dyonic RN BH, with ADM mass $M$, electric charge $Q$ and magnetic charge $P$, which was evolved using the numerical code described in \cite{sanchis2016explosion,sanchis2016dynamical,SSChargedBH1}, adapted to the power-law coupling $h(\phi)=-\alpha\phi^2$. The code uses spherical coordinates under the assumption of spherical symmetry employing the second-order Partially Implicit Runge-Kutta method developed by \cite{cordero2012partially,cordero2014partially}.

The 3+1 metric is given by $ds^2=-(\alpha_0^2+\beta^r \beta_r)dt^2+2\beta_r dtdr+e^{4\chi}\left[a\, dr^2+ b\, r^2 d\Omega_2\right]$, where the lapse $\alpha_0$, shift component $\beta^r$, and the (spatial) metric functions, $\chi,a,b$ depend on $t,r$. As in \cite{SSChargedBH1}, the scalar (initial data) perturbation to trigger the instability is a Gaussian distribution of the form
  \begin{equation} \phi=A_0e^{-(r-r_0)^2/\lambda^2} \ ,
  \end{equation} 
  with  $A_0=3\times 10^{-4}$, $r_0=10M$ and $\lambda=\sqrt{8}$. 
For the dyonic RN initial data, the conformal factor is given by
\begin{equation}
\psi  \equiv  e^{ \chi}= \biggl[\biggl(1+\frac{M}{2r}\biggl)^{2}-\frac{Q^{2}+P^{2}}{4r^{2}}\biggl]^{1/2}.
\end{equation}

At $t=0$, we choose a ``pre-collapsed" lapse $\alpha_0 = \psi^{-2}$ and a vanishing shift $\beta^{r}=0$. Initially, the electric field is given by $E^{r}=\frac{Q}{r^{2}\psi^{6}}$ and $B^{r}=\frac{P}{r^{2}\psi^{6}}$.

Since we are considering a RN BH with magnetic charge, we also have to take into account the evolution equation of the magnetic field, even in spherical symmetry. The evolution equations for the electric and magnetic fields and two extra variables, $\Psi_{\rm{E}}$ and $\Phi_{\rm{B}}$, to damp dynamically the constrains, take the following form in spherical symmetry
\begin{eqnarray}
\partial_{t}E^{r}&=&\beta^{r}\partial_{r}E^{r}-E^{r}\partial_{r}\beta^{r}+(\alpha_0 K E^{r} -D^{r}\Psi_{\rm{E}}) \ \nonumber\\
&&+2\alpha\alpha_0\,\phi\Pi B^{r}\,, \nonumber \\
\partial_{t}\Psi_{\rm{E}}&=&\alpha_0(-2\alpha \phi D_{r}\phi B^{r} - D_{i}E^{i}-\kappa_{1}\Psi_{\rm{E}})\, ,\\
\partial_{t}B^{r}&=&\beta^{r}\partial_{r}B^{r}-B^{r}\partial_{r}\beta^{r}+(\alpha_0 K B^{r} +D^{r}\Phi_{\rm{B}}) \ , \\
\partial_{t}\Phi_{\rm{B}}&=&\beta^{r}\partial_{r}\Phi_{\rm{B}}+\alpha_0(D_{i}B^{i}-\kappa_2\,\Phi_{\rm{B}}) \ ,
\end{eqnarray}
where $K$ is the trace of the extrinsic curvature $K_{ij}$, $\Pi\equiv -n^{a}\nabla_{a}\phi$, and we have chosen the damping terms $\kappa_1=\kappa_2=1$.

The Klein-Gordon equation is solved by evolving the following system of first-order equations: 
\begin{eqnarray}
\partial_{t}\phi&=&\beta^{r}\partial_{r}\phi-\alpha_0\Pi \  , \nonumber \\
\partial_{t}\Pi&=&\beta^{r}\partial_{r}\Pi+\alpha K\Pi-\frac{\alpha_0}{ae^{4\chi}}\biggl[\partial_{rr}\phi\nonumber\\
&+&\partial_{r}\phi\biggl(\frac{2}{r}-\frac{\partial_{r}a}{2a}+\frac{\partial_{r}b}{b}+2\partial_{r}\chi\biggl)\biggl]\nonumber\\
&-&\frac{\partial_{r}\phi}{ae^{4\chi}}\,\partial_{r}\alpha_0-2\alpha\alpha_0\phi\,a\,e^{4\chi}E^{r}\,B^{r} \ .
\label{eq:sist-KG}
\end{eqnarray}
\begin{figure}[ht!]
\centering
\includegraphics[width=.5\textwidth]{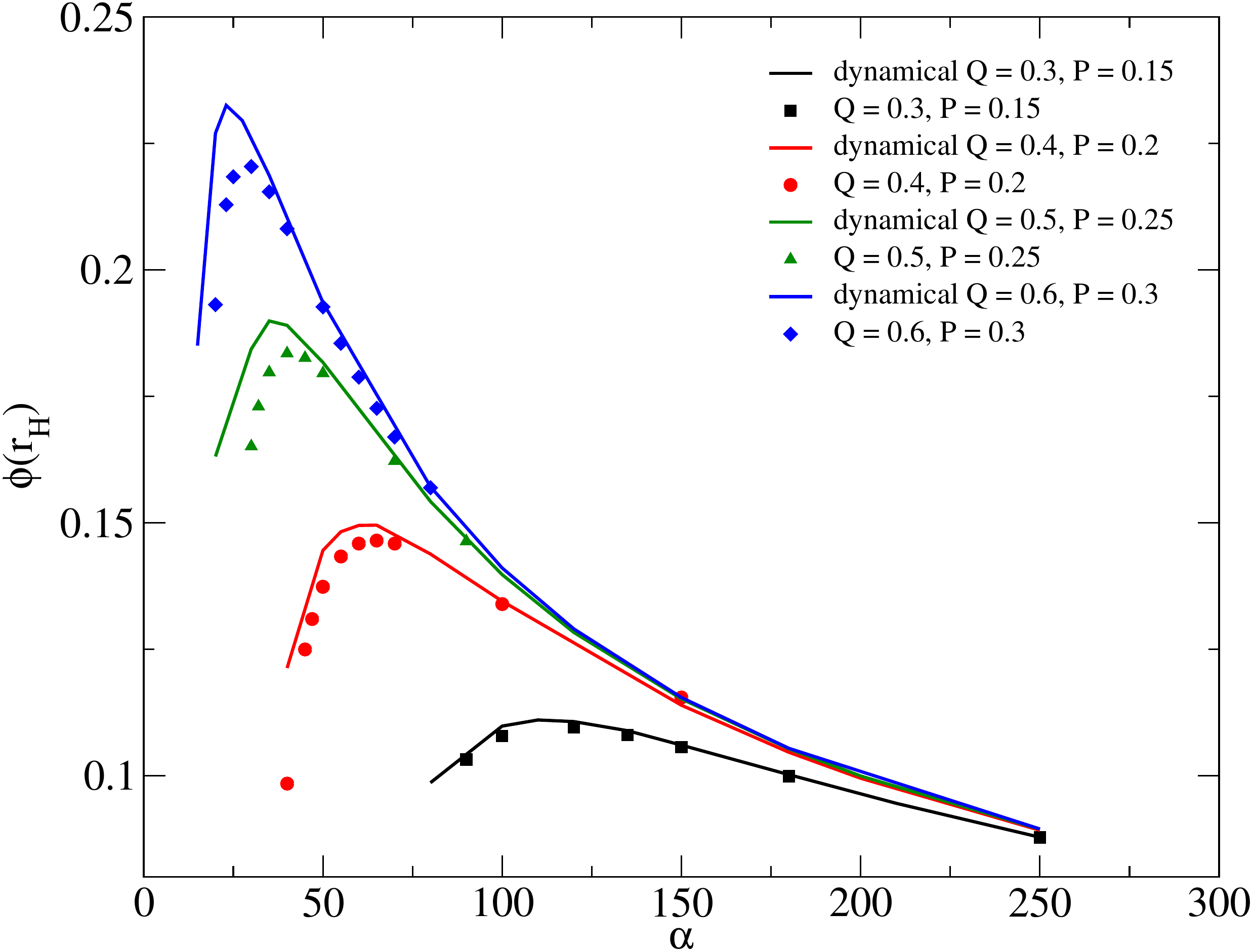}
\caption{Scalar field value at the horizon for different values of $Q$, while $\beta=0.5$, in the power-law coupling model. The solid lines are obtained from data of the scalarised BHs obtained as static solutions of the field equations. The individual points are obtained from the numerical evolutions, starting from a scalar-free RN BH with the same global charges $M,Q,P$. The agreement is better for the lower charges, showing that scalarisation only redistributes the electric charge and energy between the horizon and the scalar hair, with minor leaking towards infinity. This leaking appears to become more significant for larger charges.}
\label{fig:phih}
\end{figure}
The matter source terms in the Einstein equations are free from the axionic coupling.

The numerical simulations show that  the scalar perturbation triggers the spontaneous scalarisation of the RN BH. The horizon electric charge decreases as the energy of the field increases, while the horizon magnetic charge remains unchanged, until we reach equilibrium and a scalarised solution forms at the endpoint of the dynamical scalarisation. The scalar cloud grows near the horizon and expands radially. The radial profile of the cloud decreases monotonically with increasing radii. In Fig.~\ref{fig:phih}, we plot the scalar field value at the horizon $\phi(r_{\rm{H}})$ for both the dynamical evolutions and the static solutions with the same $Q$ and $P$. We obtain a quite good agreement with the static solutions described above. As in our previous works~\cite{SSChargedBH1,SSChargedBH2}, the dynamical solutions match better the static ones for lower values of $Q$, showing that the evolutions are more approximately conservative in that case.

%
\section{Conclusions} \label{S8}
%
\par 
This paper consider an augmented EMS sclar model and its BH solutions, in particular in the context of the spontaneous scalarisation of charged BHs. The model's novelty consists on adding an axionic-like coupling $h(\phi) F_{\mu \nu} \Tilde{F}^{\mu \nu}$ to the EMS action which has been previously studied in the context of BH spontaneous scalarisation. Depending on the choice of $h(\phi)$, the model can accommodate BHs with axionic-like and, possibly, also the standard RN BH of electrovacuum. In this case, the latter may become unstable against scalar perturbations and spontaneously scalarised, which we have shown to occur dynamically in one illustrative example.

Spontaneous scalarisation provides a dynamical mechanism for new BHs to emerge under particular circumstances. Although it is commonly considered electric charges are astrophysically irrelevant (but see, $e.g.$~\cite{EChargeImportant}), the model herein can be considered as a toy model for spontaneous scalarisation induced by higher curvature corrections, which may have astrophysical relevance. In this context, for instance, it  was pointed out recently that the observed shadow of the BH in the centre of the M87 galaxy \cite{EHT} is compatible with spontaneously scalarised Kerr BHs within some range of the parameters of the model \cite{SSKerrBHs}. However, the spontaneous scalarisation with higher curvature parity violating terms, such as the Chern-Simons term, have not yet been considered, except in the academic model in~\cite{Brihaye:2018bgc}.\footnote{Recently a study of the shadows of BHs in the EMS model has been reported~\cite{Konoplya:2019goy}.}

\medskip
%
\section*{Acknowledgements}
%
This work has been supported by Funda\c{c}\~ao para a Ci\^encia e a Tecnologia (FCT),
within project UID/MAT/04106/2019 (CIDMA), by CENTRA (FCT) strategic project UID/FIS/00099/2013, by national funds (OE), through FCT, I.P., in the scope of the framework contract foreseen in the numbers 4, 5 and 6
of the article 23, of the Decree-Law 57/2016, of August 29,
changed by Law 57/2017, of July 19. NSG  is supported by an FCT post-doctoral grant through the project PTDC/FIS-OUT/28407/2017 and A. Pombo is supported by the FCT grant PD/BD/142842/2018.   This work has further been supported by  the  European  Union's  Horizon  2020  research  and  innovation  (RISE) programmes H2020-MSCA-RISE-2015
Grant No.~StronGrHEP-690904 and H2020-MSCA-RISE-2017 Grant No.~FunFiCO-777740. The authors would like to acknowledge networking support by the
COST Action CA16104.


  \bibliographystyle{ieeetr}
  \bibliography{biblio}


\end{document}